\documentclass[showpacs,aps,prd,twocolumn]{revtex4} 
\usepackage[latin1]{inputenc}
\usepackage{pstricks}
\usepackage{amsmath}
\usepackage{float}
\usepackage{color}
\usepackage{bm}
\usepackage{amssymb}
\usepackage[dvips]{graphicx,psfrag}
\newcommand{\be}{\begin{equation}}
\newcommand{\ee}{\end{equation}}
\newcommand{\bea}{\begin{eqnarray}}
\newcommand{\eea}{\end{eqnarray}}
\newcommand{\lb}{\label}
\newcommand{\rd}{{\rm d}}

\begin{document}

\title{Chameleon dark energy models with characteristic signatures}

\author{Radouane Gannouji$^{1,2}$, Bruno Moraes$^3$, David F.~Mota$^4$,  
David Polarski$^3$, Shinji Tsujikawa$^2$,  Hans A.~Winther$^4$}

\affiliation{$^1$IUCAA, Post Bag 4, Ganeshkhind, Pune 411 007, India}

\affiliation{$^2$Department of Physics, Faculty of Science, Tokyo University of
Science, 1-3, Kagurazaka, Shinjuku-ku, Tokyo 162-8601, Japan}

\affiliation{$^3$Lab. de Physique Th\'eorique et Astroparticules, CNRS
Universit\'e Montpellier II, France}

\affiliation{$^4$Institute of Theoretical Astrophysics University of Oslo,
Norway}

\date{\today}

\begin{abstract}

In chameleon dark energy models, local gravity constraints tend to rule out 
parameters in which observable cosmological signatures can be found.  
We study viable chameleon potentials consistent 
with a number of recent observational and experimental bounds. 
A novel chameleon field potential, motivated by $f(R)$ gravity, 
is constructed where observable cosmological signatures are present 
both at the background evolution and in the growth-rate of the perturbations. 
We study the evolution of matter density perturbations on 
low redshifts for this potential and show 
that the growth index today $\gamma_0$ can have significant 
dispersion on scales relevant for large scale structures. 
The values of $\gamma_0$ can be even smaller than $0.2$
with large variations of $\gamma$ on very low redshifts
for the model parameters constrained by local gravity tests. 
This gives a possibility to clearly distinguish 
these chameleon models from the $\Lambda$-Cold-Dark-Matter
($\Lambda$CDM) model in future high-precision observations.

\end{abstract}

\pacs{04.50.Kd, 95.36.+x}

\maketitle

\section{Introduction}

The accelerated expansion of the Universe today is a very 
important challenge faced by cosmologists \cite{review}.
For an isotropic comoving perfect fluid, a substantially 
negative pressure is required to give rise to the 
cosmic acceleration.
One of the simplest candidates for dark energy is 
the cosmological constant with an equation of state 
$w_{\rm DE}=-1$,
but we generally encounter a problem
to explain its tiny energy density consistent 
with observations \cite{Weinberg}. 

There are alternative models of dark energy to the 
cosmological constant scenario. 
One of such models is quintessence based on a 
minimally coupled scalar field with a self-interacting 
potential \cite{quin}.
In order to realize the cosmic acceleration today, 
the mass of quintessence is required to be very small
($m_{\phi} \approx 10^{-33}$~GeV).
{}From a viewpoint of particle physics, such a light 
scalar field may mediate a long range force with 
standard model particles \cite{Carroll}.
For example, the string dilaton can lead
to the violation of equivalence principle through the  
coupling with baryons \cite{Gas}. 
In such cases we need to find some mechanism
to suppress the fifth force for the consistency 
with local gravity experiments.

There are several different ways to screen the field 
interaction with baryons. One is the so-called run 
away dilaton scenario \cite{Piazza} in which the field 
coupling $F(\phi)$
with the Ricci scalar $R$ is assumed to approach a constant 
value as the dilaton $\phi$ grows in time 
(e.g., $F(\phi)=C_1+C_2 e^{-\phi}$ as $\phi \to \infty$).
Another way is to consider a field potential having a 
large mass in the region of high density where 
local gravity experiments are carried out.
In this case the field does not propagate freely in the local region,
while on cosmological scales the field mass can be light enough to be
responsible for dark energy.

The latter scenario is called the chameleon mechanism in which
a density-dependent matter coupling with the field 
can allow the possibility to suppress an effective coupling 
between matter and the field outside a spherically symmetric 
body \cite{Khoury1,Khoury2}. The chameleon mechanism can be 
applied to some scalar-tensor theories such as 
$f(R)$ gravity \cite{fRlgc1,fRlgc2} and 
Brans-Dicke theory \cite{TUMTY}.
In $f(R)$ gravity, for example, there have been a number 
of viable dark energy models \cite{fRviable} 
that can satisfy both cosmological and local gravity constraints.
For such models the potential of 
an effective scalar degree of freedom 
(called ``scalaron'' \cite{star80})
in the Einstein frame is designed to have a large mass
in the region of high density.
Even with a strong coupling between the scalaron and 
the baryons ($Q=-1/\sqrt{6}$), the chameleon mechanism 
allows the $f(R)$ models to be consistent with 
local gravity constraints. 

The chameleon models are a kind of coupled quintessence 
models \cite{Amendola99} defined 
in the Einstein frame \cite{Khoury1,Khoury2}.
While the gravitational action is described by the usual 
Einstein-Hilbert action, non-relativistic matter 
components are coupled to the Einstein frame metric 
multiplied by some conformal factor which depends on a scalar
(chameleon) field. This is how the gravitational force felt 
by matter is modified. While there have been many studies
for experimental and observational aspects of the chameleon 
models \cite{Brax:2004qh}-\cite{Brax:2010kv}, 
it is not clear which chameleon potentials are viable
if the same field is to be responsible for dark energy.

In this paper we identify a number of chameleon 
potentials that can be consistent with dark energy 
as well as local gravity experiments. 
We then constrain the viable model parameter space 
by using the recent experimental and observational bounds--
such as the 2006 E\"{o}t-Wash experiment \cite{Kapner:2006si}, 
the Lunar Laser Ranging experiment \cite{Will} 
and the WMAP constraint on the time-variation of 
particle masses \cite{wmap_constraints}.
This can actually rule out some of the chameleon potentials
with natural model parameters for the matter coupling $Q$
of the order of unity.

In order to distinguish the viable chameleon dark energy 
models from the $\Lambda$CDM model, it is crucial to study 
both the modifications in the evolution of the background 
cosmology and the modified evolution of the cosmological density 
perturbations. For the former, we shall consider the evolution of the 
so-called statefinders introduced in Refs.~\cite{Sahni1,Sahni2} 
and show that these parameters can exhibit a peculiar behavior
different from those in the $\Lambda$CDM model.

On the other hand, the growth ``index'' $\gamma$ of 
matter perturbations $\delta_m$ defined through 
${\rm d} \ln \delta_m/{\rm d} \ln a =(\Omega_m^*)^{\gamma}$, 
where $a$ is a scale factor and $\Omega_m^*$ is the density 
parameter of non-relativistic matter, 
is an important quantity 
that allows to discriminate between 
different dark energy models and interest in this quantity 
was revived in the context of dark energy models 
\cite{Stein,Linder}. 

Its main importance for the study of dark energy models stems 
from the fact that for $\Lambda$CDM the quantity $\gamma$ is known 
to be nearly constant with respect to the redshift $z$, i.e.
$\gamma_{\Lambda {\rm CDM}}= \gamma_0 - 0.02 z$ 
to exquisite accuracy \cite{PG07}, 
with $\gamma_0\equiv \gamma(z=0)\approx 0.555$ \cite{Stein}.
As emphasized in Ref.~\cite{PG07}, large variations of $\gamma$ on 
low redshifts could signal that we are dealing with a dark energy 
model outside General Relativity. This was indeed found for some 
scalar-tensor dark energy models \cite{GP08} and $f(R)$ models 
\cite{GMP08,MSY10,Narikawa}. Such large variations can also 
occur in models where dark energy interacts with matter \cite{ASS09,HWJ09}. 
This is exactly the case in chameleon models, which we investigate in 
this paper, because of the direct coupling between the chameleon field 
$\phi$ and all dust-like matter. This direct coupling is however not confined 
to the dark sector as in standard coupled quintessence.

An additional important point is the possible appearance of 
a scale-dependence or dispersion in $\gamma$. 
Hence the behavior of $\gamma$ on low redshifts can be both 
time-dependent and scale-dependent \cite{GMP09,Tsujikawa:2009ku,BT10}. 
This dispersion can also be present in the models investigated here. 
Using the observations of large scale structure
and weak lensing surveys, one can hope to detect such peculiar 
behaviors of $\gamma$ (see e.g., Ref.~\cite{Euclid}). If this is 
the case this would signal that the gravitational law may be 
modified on scales relevant to large scale structures 
\cite{Linder,Gannouji,Luca,Tsujikawa:2009ku,Koivisto,Nesseris,BT10,SD09,Mota:2007zn}.

In this paper we study the evolution of $\gamma$ as well 
as its dispersion, its dependence on the wavenumbers of 
perturbations. 
We shall show that some of the chameleon models 
investigated here can be clearly distinguished from $\Lambda$CDM 
through the behavior of $\gamma$ exhibiting 
both large variations and significant dispersion, with the possibility 
to obtain small values of $\gamma$ today as low as $\gamma_0 \lesssim 0.2$.

\section{Chameleon cosmology}
\label{cosmologysec}

%
\subsection{Background equations}

In this section we review the basic background evolution of 
chameleon cosmology. 
We consider the chameleon theory described 
by the following action \cite{Khoury1,Khoury2,EP00}
\bea
\mathcal{S}\,&=&\,\int{{\rm d}^4x}\sqrt{-g}\left[ 
\frac{R}{16\pi G} -\frac{1}{2} g^{\mu\nu} 
\partial_{\mu}\phi\ \partial_{\nu}\phi 
- V(\phi) \right] \nonumber \\ 
&& +\mathcal{S}_m \left[ \Psi_m; A^2(\phi)\,
g_{\mu\nu} \right]\,,
\lb{action}
\eea
where $g$ is the determinant of the (Einstein frame) metric $g_{\mu \nu}$, 
$R$ is the Ricci scalar, $G$ is the bare gravitational constant, 
$\phi$ is a scalar field with a potential $V(\phi)$, and 
$\mathcal{S}_m$ is the matter action with matter fields $\Psi_m$.
At low redshifts it is sufficient to consider only 
non-relativistic matter (cold dark matter and baryons),
but for a general dynamical analysis 
including high redshifts radiation must be included. 

We assume that non-relativistic matter is universally 
coupled to the (Jordan frame) metric $A^2(\phi)~g_{\mu \nu}$, the 
Einstein frame metric $g_{\mu \nu}$ multiplied by a 
field-dependent (conformal) factor $A^2 (\phi)$. This direct coupling 
to the field $\phi$ is how the gravitational interaction is modified. 
We can generalize this to arbitrary functions $A^{(i)}(\phi)$ for 
each matter component $\rho_i$, but in this work we will take the same 
function $A(\phi)$ for all components. 
We write the function $A(\phi)$ in the form 
\be
A (\phi)=e^{Q \phi/M_{\rm pl}}\,,
\label{Aphi}
\ee
where $M_{\rm pl}=1/\sqrt{8\pi G}$ is the reduced Planck mass
and $Q$ describes the strength of the coupling between the field 
$\phi$ and non-relativistic matter.
In the following we shall consider the case in which $Q$ is constant.
In fact, the constant coupling arises for Brans-Dicke theory
by a conformal transformation to the Einstein 
frame \cite{Khoury2,TUMTY}.
Even when $|Q|$ is of the order of unity, it is possible to make
the effective coupling between the field and matter small 
through the chameleon mechanism.

Let us consider the scalar field $\phi$ together with 
non-relativistic matter (density $\rho^*_m$) and 
radiation (density $\rho_r$) in a spatially flat 
Friedmann-Lema\^{\i}tre-Robertson-Walker
(FLRW) space-time with a time-dependent scale 
factor $a(t)$ and a metric 
\be
{\rm d}s^{2}= g_{\mu\nu}\,{\rm d}x^{\mu}\,{\rm d}x^{\nu} 
= -{\rm d}t^{2}+a^{2}(t)\,{\rm d}{\bm x}^{2}\,.
\ee
The corresponding background equations are given by
\bea
 &  & 3H^2=\left( \rho_{\phi}
 +\rho_m^*+\rho_r \right)/M_{\rm pl}^2\,,
 \label{cosmo1} \\
 &  & \ddot{\phi}+3H\dot{\phi}+V_{,\phi}=-Q\rho_m^*/M_{\rm pl}\,,
\label{cosmo2} \\
 &  & \dot{\rho}_{m}^*+3H\rho_{m}^*=Q\rho_{m}^*
 \dot{\phi}/M_{\rm pl}\,,
\label{cosmo3} \\
 &  & \dot{\rho}_{r}+4H\rho_{r}=0\,,
 \label{cosmo4}
\eea
where $\rho_{\phi} \equiv \dot{\phi}^2/2+V(\phi)$, 
$V_{,\phi} \equiv {\rm d}V/{\rm d}\phi$, and 
a dot represents a derivative with respect to 
cosmic time $t$.
The quantity $\rho^*_m$ is the energy density of 
non-relativistic 
matter in the Einstein frame and we have kept 
the star to avoid any confusion. 
Integration of Eq.~(\ref{cosmo3}) gives the solution 
$\rho_m^* \propto a^{-3} e^{Q \phi/M_{\rm pl}}$.
We define the conserved matter density:
\be
\rho_m \equiv e^{-Q \phi/M_{\rm pl}} \rho_m^*\,,
\ee
which satisfies the standard continuity equation, 
$\dot{\rho}_m+3H\rho_m=0$.
Then the field equation (\ref{cosmo2}) can be written 
in the form 
\be
\ddot{\phi}+3H \dot{\phi}+V_{{\rm eff},\phi}=0\,,
\ee
where $V_{\rm eff}$ is the effective potential defined by 
\be
V_{\rm eff} \equiv V(\phi)+e^{Q \phi/M_{\rm pl}} \rho_m\,.
\label{Veff}
\ee
We emphasize that it is the Einstein frame which is the physical 
frame. Due to the coupling between the field and matter, 
particle masses do evolve with time in our model.

We consider runaway positive potentials $V(\phi)$ in the region 
$\phi>0$, which monotonically decrease and have a positive mass 
squared, i.e. $V_{,\phi}<0$ and $V_{,\phi \phi}>0$.
We also demand the following conditions
\be
\lim_{\phi \to 0} \left| \frac{V_{,\phi}}{V} \right|
=\infty\,,\qquad
\lim_{\phi \to \infty} \frac{V_{,\phi}}{V}=0\,.
\label{lamre}
\ee
The former is required to have a large mass 
in the region of high density, 
whereas we need the latter condition to realize the late-time
cosmic acceleration in the region of low density.
At $\phi=0$ the potential approaches either $\infty$ or a finite 
positive value $V_0$. In the limit $\phi \to \infty$ we have 
either $V \to 0$ or $V \to V_{\infty}$, where $V_{\infty}$
is a nonzero positive constant.
If $Q>0$ the effective potential $V_{\rm eff}(\phi)$
has a minimum at the field value $\phi_{m}~(>0)$ satisfying 
the condition $V_{{\rm eff},\phi} (\phi_m)=0$, i.e.
\be
V_{,\phi} (\phi_m)+Q (\rho_m/M_{\rm pl})
e^{Q\phi_m/M_{\rm pl}}=0\,.
\label{extremum}
\ee

If the potential satisfies the conditions 
$V_{,\phi}>0$ and $V_{,\phi \phi}>0$ in the region $\phi<0$, 
there exists a minimum at $\phi=\phi_m~(<0)$ provided 
that $Q<0$. In fact this situation arises in the context of 
$f(R)$ dark energy models \cite{fRlgc1,fRlgc2}.
Since the analysis in the latter is equivalent to that 
in the former, we shall focus on 
the case $Q>0$ and $V_{,\phi}<0$
in the following discussion. 

\subsection{Dynamical system}

In order to discuss cosmological dynamics, it is convenient to 
introduce the following dimensionless variables
\be
x_{1}\equiv\frac{\dot{\phi}}{\sqrt{6}HM_{\rm pl}}\,,\quad 
x_{2}\equiv\frac{\sqrt{V}}{\sqrt{3}HM_{\rm pl}}\,,\quad
x_{3}\equiv\frac{\sqrt{\rho_{r}}}{\sqrt{3}HM_{\rm pl}}\,.
\ee
Equation (\ref{cosmo1}) expresses the constraint existing 
between these variables, i.e.
\be
\Omega_m^* \equiv \frac{\rho_m^*}{3H^2 M_{\rm pl}^2}=
1-\Omega_{\phi}-\Omega_{r}\,,
\ee
where 
\be
\Omega_{\phi} \equiv x_1^2+x_2^2\,,\qquad
\Omega_{r} \equiv x_3^2\,.
\label{Omephi}
\ee
Taking the time-derivative of Eq.~(\ref{cosmo1})
and making use of Eqs.~(\ref{cosmo2})-(\ref{cosmo4}), 
it is straightforward to derive the following equation 
\be
\frac{H'}{H}=-\frac{1}{2}\left(3+3x_{1}^{2}-3x_{2}^{2}
+x_{3}^{2}\right)\,,\label{dotHcou}
\ee
where a prime represents a derivative with respect to $N\equiv \ln a$. 
A useful quantity is the effective equation of state
\be
w_{{\rm eff}} = -1 - \frac{2}{3} \frac{H'}{H} = 
x_{1}^{2}-x_{2}^{2}+\frac{x_3^2}{3}\,.
\ee

We also introduce the field equation of state $w_{\phi}$, as
\be
w_{\phi} \equiv 
\frac{\dot{\phi}^2/2-V(\phi)}{\dot{\phi}^2/2+V(\phi)} = 
\frac{x_{1}^{2}-x_{2}^{2}}{x_{1}^{2}+x_{2}^{2}}\,.
\label{wphi}
\ee
Using Eqs.~(\ref{cosmo1})-(\ref{cosmo4}), we obtain the following equations
\bea
&  & x_{1}'=-3x_{1}+\frac{\sqrt{6}}{2}\lambda
x_{2}^{2}-x_{1}\frac{H'}{H} \nonumber \\
&
&~~~~~~-\frac{\sqrt{6}}{2}Q\left(1-x_{1}^{2}-x_{2}^{2}-x_{3}^{2}\right)\,,\label
{x1cou}\\
 &  & x_{2}'=-\frac{\sqrt{6}}{2}\lambda
x_{1}x_{2}-x_{2}\frac{H'}{H}\,,\label{x2cou}\\
 &  & x_{3}'=-2x_{3}-x_{3}\frac{H'}{H}\,,\label{x3cou}\\
 &  & \lambda'=-\sqrt{6}\lambda^2 (\Gamma-1)x_1\,,
 \label{lambdacou}
\eea
where 
\be
\lambda \equiv -\frac{M_{\rm pl}V_{,\phi}}{V}\,,\qquad
\Gamma \equiv \frac{VV_{,\phi \phi}}{V_{,\phi}^2}\,.
\ee
{}From the conditions (\ref{lamre}) it follows that 
the quantity $\lambda$ decreases from $\infty$ to 0 as $\phi$
grows from 0 to $\infty$. Since $x_1>0$ in Eq.~(\ref{lambdacou}), 
the condition $\lambda'<0$ translates into
\be
\Gamma=\frac{VV_{,\phi \phi}}{V_{,\phi}^2}>1\,.
\label{Gamcon}
\ee
Chameleon potentials shallower than the exponential potential 
($\Gamma=1$) can satisfy this condition.

Once the field settles down at the minimum of the effective potential
(\ref{Veff}), we have 
\be
x_1 \simeq 0\,, \qquad 
x_2 \simeq \sqrt{\frac{Q}{\lambda} \Omega_m^*}\,,
\label{ins}
\ee
which gives $w_{\phi} \simeq -1$ from 
Eq.~(\ref{wphi}). As the matter density $\rho^*_m$ decreases,  
the field evolves slowly along the instantaneous minima characterized by 
(\ref{ins}). We require that $\lambda \gg Q={\cal O}(1)$ during radiation 
and deep matter eras for consistency with local gravity constraints 
in the region of high density.
For the dynamical system (\ref{x1cou})-(\ref{lambdacou}) there is 
another fixed point called the ``$\phi$-matter-dominated era 
($\phi$MDE)'' \cite{Amendola99} 
where $\Omega_{\phi}=w_{\rm eff}=2Q^2/3$.
However, since we are considering the case in which $Q$ is 
of the order of unity, the effective equation of state $w_{\rm eff}$
is too large to be compatible with observations.
Only when $Q<{\cal O}(0.1)$ the $\phi$MDE can be 
responsible for the matter era \cite{Amendola99}.

When the chameleon is slow-rolling along the minimum, 
we obtain the following relation from 
Eqs.~(\ref{Omephi}) and (\ref{ins}):
\be
\frac{\lambda}{Q} \simeq \frac{\Omega_m^*}
{\Omega_{\phi}}\,.
\label{cosmocon}
\ee
While $\lambda \gg Q$ during the radiation and matter eras, 
$\lambda$ becomes the same order as $Q$ around the 
present epoch.
The field potential is the dominant contribution on the r.h.s. of 
Eq.~(\ref{cosmo1}) today, so that 
\be
V(\phi_0) \simeq 3H_0^2M_{\rm pl}^2 
\simeq \rho_c\,,
\label{enecon}
\ee
where the subscript ``0'' represents present values and 
$\rho_c \simeq 10^{-29}$\,g/cm$^3$ 
is the critical density today.

\section{Chameleon mechanism}

In this section we review the chameleon mechanism as a way 
to escape local gravity constraints. 
In addition to the cosmological constraints discussed in the 
previous section, this will enable us to restrict the forms 
of chameleon potentials.

Let us consider a spherically symmetric space-time in the weak 
gravitational background with the neglect of the backreaction 
of metric perturbations.
As in the previous section we consider the case in which 
couplings $Q_{i}$ are the same for each matter 
component ($Q_i=Q$), i.e., 
in which the function $A(\phi)$ is given by Eq.~(\ref{Aphi}). 
Varying the action (\ref{action}) with respect to $\phi$
in the Minkowski background, we obtain the field equation  
\be
\frac{\rd^{2}\phi}{\rd r^{2}}+\frac{2}{r}\frac{\rd\phi}{\rd r}
=\frac{\rd V_{{\rm eff}}}{\rd\phi}\,,
\label{sphequa}
\ee
where $r$ is the distance from the center of symmetry and 
$V_{\rm eff}$ is defined in Eq.~(\ref{Veff}).

Assuming that a spherically symmetric object (radius $r_c$ and mass $M_c$) 
has a constant density $\rho_m=\rho_{A}$ with a homogeneous density 
$\rho_m=\rho_{B}$ outside the body, the effective potential has two minima
at $\phi=\phi_A$ and $\phi=\phi_B$ satisfying the conditions
\bea
V_{,\phi} (\phi_A)&+&Q(\rho_A/M_{\rm pl})e^{Q \phi_A/M_{\rm pl}}=0\,,\\
V_{,\phi} (\phi_B)&+&Q(\rho_B/M_{\rm pl})e^{Q \phi_B/M_{\rm pl}}=0\,.
\eea
Since $Q\phi_A/M_{\rm pl} \ll 1$ and $Q\phi_B/M_{\rm pl} \ll 1$
for viable field potentials in the regions of high density, 
the conserved matter density $\rho_m$ is practically 
indistinguishable from the matter density $\rho_m^*$ 
in the Einstein frame. 

The field profile inside and outside the body can be found analytically.
Originally this was derived in Refs.~\cite{Khoury1,Khoury2} 
under the assumption that the field is frozen around $\phi=\phi_A$ 
in the region $0<r<r_1$, where $r_1~(<r_c)$
is the distance at which the field begins to evolve. 
It is possible, even without this assumption, to derive analytic solutions 
by considering boundary conditions at the center of 
the body \cite{Tamaki:2008mf}.

We consider the case in which the mass squared 
$m_B^2 \equiv \frac{\rd^2 V_{\rm eff}}{\rd \phi^2} (\phi_B)$ 
outside the body satisfies the condition $m_B r_c \ll 1$, so that the
$m_B$-dependent terms can be negligible when we match solutions at $r=r_c$.
The resulting field profile outside the body ($r>r_c$) 
is given by \cite{Tamaki:2008mf}
\be
\phi(r)=\phi_{B}-\frac{Q_{{\rm eff}}}{4\pi M_{\rm pl}}\frac{M_{c}}{r}\,,
\label{spheq}
\ee
where the effective coupling $Q_{\rm eff}$ between the field and matter is
\bea
Q_{{\rm eff}}&=&Q\biggl[1-\frac{r_{1}^{3}}{r_{c}^{3}}+
3\frac{r_{1}}{r_{c}}\frac{1}{(m_{A}r_{c})^{2}} \nonumber \\
&&~~~\times \left\{ \frac{m_{A}r_{1}(e^{m_{A}r_{1}}+
e^{-m_{A}r_{1}})}{e^{m_{A}r_{1}}-e^{-m_{A}r_{1}}}-1\right\} \biggr]\,.
\label{Qeff}
\eea
The mass $m_A$ is defined by 
$m_A^2 \equiv \frac{\rd^2 V_{\rm eff}}{\rd \phi^2}
(\phi_A)$. 
The distance $r_{1}$ is determined by the condition 
$m_{A}^{2}\left[\phi(r_{1})-\phi_{A}\right]=Q\rho_{A}$,
which translates into 
\bea
& &\phi_{B}-\phi_{A}+Q\rho_{A}(r_{1}^{2}-r_{c}^{2})/(2M_{\rm pl}) \nonumber \\
&&=\frac{6QM_{\rm pl}\Phi_{c}}{(m_{A}r_{c})^{2}}
\frac{m_{A}r_{1}(e^{m_{A}r_{1}}+e^{-m_{A}r_{1}})}{e^{m_{A}r_{1}}-e^{-m_{A}r_{1}}
}\,,\label{r1con}
\eea
where $\Phi_{c}=M_{c}/(8\pi r_{c})=\rho_{A}r_{c}^{2}/(6M_{\rm pl}^2)$ 
is the gravitational potential at the surface of the body. 

The fifth force exerting on a test particle of a unit mass and a coupling $Q$
is given by ${\bm F}=-Q \nabla \phi/M_{\rm pl}$.
Using Eq.~(\ref{spheq}), the amplitude of the fifth force
in the region $r>r_{c}$ is 
\be
F=2\left|QQ_{{\rm eff}}\right|\frac{G M_{c}}{r^{2}}\,.
\ee
As long as $|Q_{{\rm eff}}| \ll 1$, it is possible to make the fifth
force suppressed relative to the gravitational force $G M_{c}/r^{2}$.
{}From Eq.~(\ref{Qeff}) the effective coupling $Q_{\rm eff}$
can be made much smaller than $Q$ provided that the conditions
$\Delta r_{c}\equiv r_{c}-r_{1}\ll r_{c}$ and $m_{A}r_{c}\gg1$ are
satisfied. Hence we require that the body has a thin-shell and that
the field is heavy inside the body for the chameleon mechanism to work.

When the body has a thin-shell ($\Delta r_{c}\ll r_{c}$), one can
expand Eq.~(\ref{r1con}) in terms of the small parameters 
$\Delta r_{c}/r_{c}$ and $1/(m_{A}r_{c})$. This leads to 
\be
\epsilon_{{\rm th}}\equiv\frac{\phi_{B}-\phi_{A}}{6QM_{\rm pl} \Phi_{c}}
\simeq \frac{\Delta r_{c}}{r_{c}}+\frac{1}{m_{A}r_{c}}\,,
\label{epth}
\ee
where $\epsilon_{{\rm th}}$ is called the thin-shell parameter.
As long as $m_{A}r_{c}\gg(\Delta r_{c}/r_{c})^{-1}$, this recovers
the relation $\epsilon_{{\rm th}}\simeq\Delta r_{c}/r_{c}$
\cite{Khoury1,Khoury2}.
The effective coupling (\ref{Qeff}) is approximately given by 
\be
Q_{{\rm eff}} \simeq 
3Q\epsilon_{{\rm th}}\,.\label{Qeff2}
\ee
If $\epsilon_{{\rm th}}$ is much smaller than 1 
then one has $Q_{{\rm eff}} \ll Q$, so that the models can be 
consistent with local gravity constraints. 

As an example, let us consider the experimental bound that 
comes from the solar system tests of 
the equivalence principle, namely the Lunar Laser Ranging (LLR) experiment, 
using the free-fall acceleration of the Moon ($a_{{\rm Moon}}$) 
and the Earth ($a_{\oplus}$)
toward the Sun (mass $M_{\odot}$) \cite{Khoury2,Tamaki:2008mf,fRlgc2}. 
The experimental bound on the difference
of two accelerations is given by 
\be
\frac{2|a_{{\rm Moon}}-a_{\oplus}|}
{(a_{{\rm Moon}}+a_{\oplus})}<10^{-13}\,.
\label{etamoon}
\ee

Under the conditions that the Earth, the Sun, and the Moon have thin-shells,
the field profiles outside the bodies are given as in Eq.~(\ref{spheq})
with the replacement of corresponding quantities. The acceleration 
induced by a fifth force with the field profile $\phi(r)$
and the effective coupling $Q_{{\rm eff}}$ is
$a^{{\rm fifth}}=|Q_{{\rm eff}}\nabla\phi(r)/M_{\rm pl}|$.
Using the thin-shell parameter $\epsilon_{{\rm th},\oplus}$ for the
Earth, the accelerations $a_{\oplus}$ and $a_{{\rm Moon}}$ 
are \cite{Khoury2} 
\bea
a_{\oplus} & \simeq & \frac{G M_{\odot}}{r^{2}}\left[1+
18Q^{2}\epsilon_{{\rm
th},\oplus}^{2}\frac{\Phi_{\oplus}}{\Phi_{\odot}}\right]\,,\\
a_{{\rm Moon}} & \simeq & \frac{G M_{\odot}}{r^{2}}\left[1+18Q^{2}\epsilon_{{\rm
th},
\oplus}^{2}\frac{\Phi_{\oplus}^{2}}{\Phi_{\odot}\Phi_{{\rm Moon}}}\right]\,,
\eea
where $\Phi_{\odot}\simeq2.1\times10^{-6}$,
$\Phi_{\oplus}\simeq7.0\times10^{-10}$,
and $\Phi_{{\rm Moon}}\simeq3.1\times10^{-11}$ are the gravitational
potentials of Sun, Earth and Moon, respectively. 
Then the condition (\ref{etamoon}) reads
\be
\epsilon_{{\rm th},\oplus}<\frac{8.8\times10^{-7}}{Q}\,.\label{boep}
\ee
Using the value $\Phi_{\oplus}\simeq7.0\times10^{-10}$, 
the bound (\ref{boep}) translates into 
\be
\phi_{B,\oplus} \lesssim 10^{-15}M_{\rm pl}\,,
\label{bou2}
\ee
where we used the condition $\phi_{B,\oplus} \gg \phi_{A,\oplus}$.
For the Earth one has $\rho_{A}\simeq 5$\,g/cm$^{3} 
({\rm mean~density~of~the~Earth}) 
 \gg \rho_{B}\simeq10^{-24}$\,g/cm$^{3}$ 
(dark matter/baryon density in our galaxy), so that 
the condition $\phi_{B,\oplus} \gg \phi_{A,\oplus}$ is satisfied.

In Sec.~\ref{viablesec} we constrain viable chameleon potentials
by employing the condition (\ref{bou2}) together with 
the cosmological condition we discussed in Sec.~\ref{cosmologysec}.
In Sec.~\ref{lgcsec} we restrict the allowed model parameter space 
further by using a number of recent local gravity and 
observational constraints.

\section{Viable chameleon potentials}
\label{viablesec}

We now discuss the forms of viable field potentials that can be 
in principle consistent with both local gravity and 
cosmological constraints.
Let us consider the potential 
\be
V(\phi)=M^4 f(\phi)\,,
\ee
where $M$ is a mass scale and $f(\phi)$ is a dimensionless 
function in terms of $\phi$.

The local gravity constraint coming from the LLR experiment
is given by Eq.~(\ref{bou2}), 
where $\phi_{B,\oplus}$ is determined by solving 
\be
\left| M_{\rm pl}f_{,\phi}(\phi_{B,\oplus}) \right| 
\simeq Q\rho_B/M^4\,.
\label{con2}
\ee
Here we take $\rho_B \simeq 10^{-24}$\,g/cm$^3$ 
for the homogeneous density outside the Earth.
Once the form of $f(\phi)$ is specified, the constraint 
on the model parameter, e.g., $M$, can be derived.

{}From the cosmological constraint (\ref{cosmocon}), 
$\lambda/Q$ is of the order of 1 today. 
Then it follows that 
\bea
\label{con3}
\left| M_{\rm pl}f_{,\phi}(\phi_0) \right| \simeq
Q f(\phi_0) \simeq Q\rho_c/M^4\,,
\eea
where we used Eq.~(\ref{enecon}).

We also require the condition (\ref{Gamcon}), i.e.
\be
\Gamma =\frac{ff_{,\phi \phi}}{f_{,\phi}^2}>1\,,
\label{con4}
\ee
for all positive values of $\phi$.
We shall proceed to find viable potentials satisfying 
the conditions (\ref{bou2}), (\ref{con2}), (\ref{con3}), 
and (\ref{con4}).
{}From Eqs.~(\ref{con2}) and (\ref{con3})
we obtain
\be
\frac{f_{,\phi} (\phi_{B,\oplus})}{f_{,\phi}(\phi_0)}
\simeq \frac{\rho_B}{\rho_c} \simeq 10^{5}\,.
\label{con5}
\ee

Let us consider the inverse power-law potential 
$V(\phi)=M^{4+n}\phi^{-n}$ ($n>0$), i.e.
\be
f(\phi)=(M/\phi)^{n}\,.
\ee
Since $\Gamma=(n+1)/n>1$, the condition (\ref{con4}) 
is automatically satisfied.
The cosmological constraint (\ref{con3}) gives
\be
\phi_0 \simeq nM_{\rm pl}/Q\,.
\label{conphi0}
\ee
{}From Eq.~(\ref{con5}) we find the relation 
between $\phi_0$ and $\phi_{B,\oplus}$\,:
\be
\phi_0 \simeq 10^{5/(n+1)} \phi_{B,\oplus}\,.
\label{incon1}
\ee
Using the LLR bound (\ref{bou2}), it follows that 
\be
\phi_0 \lesssim 10^{-\frac{5(3n+2)}{n+1}}M_{\rm pl}\,.
\label{incon2}
\ee
This is incompatible with the cosmological constraint (\ref{conphi0})
for $n \ge 1$ and $Q={\cal O}(1)$.
Hence the inverse power-law potential is not viable.

\subsection{Inverse power-law potential $+$ constant}

The reason why the inverse power-law potential does not work is that 
the field value today required for cosmic acceleration is of the order
of $M_{\rm pl}$, while the local gravity constraint demands a much 
smaller value.
This problem can be circumvented by taking into account a constant 
term to the inverse power-law potential.
Let us then consider the potential 
$V(\phi)=M^4 \left[1+\mu (M/\phi)^n \right]$ ($n>0$), i.e.
\be
f(\phi)=1+\mu (M/\phi)^n\,,
\label{fphi}
\ee
where $\mu$ is a positive constant.
The rescaling of the mass term $M$ always allows 
to normalize the constant to be unity in Eq.~(\ref{fphi}). 
For this potential the quantity $\Gamma$ reads
\be
\Gamma=\frac{n+1}{n} \left[ 1+\frac{1}{\mu}
\left( \frac{\phi}{M} \right)^n \right]\,,
\ee
which satisfies the condition $\Gamma>1$.
In the region $\mu (M/\phi)^n \gg 1$ we have that 
$\Gamma \simeq (n+1)/n$, which recovers the case of the 
inverse power-law potential.
Meanwhile, in the region  $\mu (M/\phi)^n \ll 1$, one has 
$\Gamma \gg 1$. The latter property comes from the fact 
that the potential becomes shallower as the field $\phi$
increases. This modification of the potential allows a 
possibility that the model can be consistent with 
both cosmological and local gravity constraints.

The addition of a constant term to the inverse power-law potential
does not affect the condition (\ref{con5}), which means that 
the resulting bounds (\ref{incon1}) and (\ref{incon2}) 
are not subject to change. 
On the other hand, the cosmological constraint (\ref{conphi0}) 
is modified. Let us consider the case where the condition 
$\mu (M/\phi_0)^n \ll 1$ is satisfied today, i.e. 
$f(\phi_0) \simeq 1$. 
{}From Eq.~(\ref{con3}) it follows that
\be
M \simeq \rho_c^{1/4} \simeq 10^{-12}\,{\rm GeV}\,,
\label{Mcon}
\ee
and 
\be
\phi_0 \simeq \left( n\mu \frac{M^n}{M_{\rm pl}^n} 
\right)^{1/(n+1)}M_{\rm pl} 
\simeq ( 10^{-30n}n \mu)^{1/(n+1)}M_{\rm pl}\,.
\label{inppo}
\ee
Hence the field value $\phi_0$ today can be much smaller than 
the Planck mass, unlike the inverse power-law potential. 
{}From Eqs.~(\ref{incon2}) and (\ref{inppo}) we get the constraint
\be
\mu \lesssim 10^{15n-10}/n\,.
\label{mucon}
\ee

If $n=1$, for example, one has $\mu \lesssim 10^5$.
For larger $n$ the bound on $\mu$ becomes even weaker.
We note that the condition $\mu(M/\phi)^n<1$ is satisfied
for $\phi/M_{\rm pl} \gtrsim 10^{-10/n-15}/n^{1/n}$.
This shows that even the field value such as 
$\phi_{B,\oplus}=10^{-15}M_{\rm pl}$ satisfies the
condition $\mu(M/\phi_{B,\oplus})^n<1$.
Thus the term $\mu (M/\phi)^n$ is smaller than 1
for the field values we are interested in 
($\phi_{B,\oplus} \lesssim \phi \lesssim \phi_0$).

A large range of experimental bounds for this model 
has been derived in the literature, see
Refs.~\cite{Khoury2,Mota:2006fz,Mota:2006ed,Brax:2007vm}. 
For $Q=n=1$, it was found in Ref.~\cite{Mota:2006fz} that the model 
is ruled out by the E\"{o}t-Wash experiment unless 
$\mu \lesssim 10^{-5}$. This applies for general $n$: to obtain a viable model for
$Q$ of the order of unity one must impose a fine-tuning 
$n\gg 1$ or $\mu \ll 1$. 

The potentials, which have only one mass scale equivalent to the dark 
energy scale, are usually strongly constrained by the E\"{o}t-Wash experiment.
We shall look into this issue in more details in Sec.~\ref{lgcsec}.

\subsection{Construction of viable chameleon potentials relevant to dark energy}

The discussion given above shows that a function  
$f(\phi)$ that monotonically decreases without a constant term 
is difficult to satisfy both cosmological and local gravity constraints.
This is associated with the fact that for any power-law form of $f(\phi)$
the condition $|M_{\rm pl}f_{,\phi}(\phi_0)/f(\phi_0)| \simeq 1$ leads 
to the overall scaling of the function $f(\phi_0)$ itself, giving $\phi_0$
of the order of $M_{\rm pl}$.
The dominance of a constant term in $f(\phi_0)$ changes this situation, 
which allows a much smaller value of $\phi_0$ relative to $M_{\rm pl}$.

Another example similar to $V(\phi)=M^4 [1+\mu (M/\phi)^n]$
is the potential \cite{Brax:2004qh}
\be
V(\phi)=M^4 \exp[\mu (M/\phi)^n]\,,
\label{vipo1}
\ee
where $\mu>0$ and $n>0$.
For this model the quantity
\be
\Gamma=1+\frac{n+1}{n}\frac{1}{\mu} 
\left( \frac{\phi}{M} \right)^n\,,
\ee
is larger than 1.
In the asymptotic regimes characterized by $\mu (M/\phi)^n \gg 1$ and 
$\mu (M/\phi)^n \ll 1$ we have $\Gamma \simeq 1$
and $\Gamma \gg 1$, respectively. 
When $\mu (M/\phi)^n \ll 1$ the function $f(\phi)=\exp[\mu (M/\phi)^n]$ 
can be approximated as $f(\phi) \simeq 1+\mu (M/\phi)^n$, which corresponds to
Eq.~(\ref{fphi}). In this case the constraints on the model parameters
are the same as those given in Eqs.~(\ref{Mcon})-(\ref{mucon}).

There is another class of potentials that behaves as 
$V(\phi) \simeq M^4 [1-\mu (\phi/M)^n]$ ($\mu>0, 0<n<1$)
in the region $\mu (\phi/M)^n \ll 1$.
In fact this asymptotic form corresponds to the potential that 
appears in $f(R)$ dark energy models.
While the potential is finite at $\phi=0$ the derivative $|V_{,\phi}|$
diverges as $\phi \to 0$ for $0<n<1$, 
so that the first of the condition (\ref{lamre})
is satisfied. In order to keep the potential positive 
we need some modification of $V$ in the region $\mu (\phi/M)^n>1$.

In scalar-tensor theory it was shown in Ref.~\cite{TUMTY} that 
the Jordan frame potential of the form 
$U(\phi)=M^4 [1-\mu (1-e^{-2Q\phi/M_{\rm pl}})^n]$
($0<\mu<1$, $0<n<1$) can  satisfy both cosmological and local gravity 
constraints. In this case the potential $V(\phi)$ in the Einstein frame 
is given by $V(\phi)=e^{4Q\phi/M_{\rm pl}}U(\phi)$, which possesses 
a de Sitter minimum due to the presence of the conformal factor. 
Cosmologically the solutions finally approach the
de Sitter fixed point, so that the late-time cosmic 
acceleration can be realized.

\begin{figure}
\begin{centering}
\includegraphics[height=3.0in,width=3.0in]{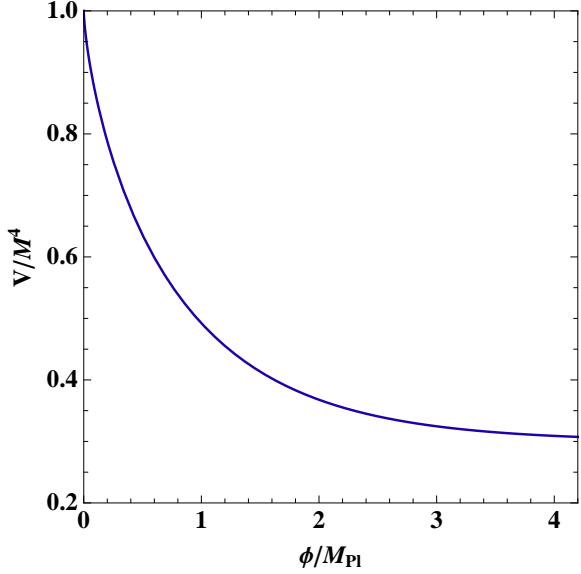}
\par\end{centering}
\caption{The potential (\ref{vipo2}) versus the field $\phi$
for $\mu=0.7$ and $n=0.7$.
The potential has a finite value $V=M^4$ at $\phi=0$, 
but its derivatives diverge ($|V_{,\phi}| \to \infty$ and 
$V_{,\phi \phi} \to \infty$) as $\phi \to 0$.}
\label{potential}
\end{figure}

Now we would like to consider a runaway positive potential 
in the Einstein frame. One example is 
\be
V(\phi)= M^4 [ 1-\mu (1-e^{-\phi/M_{\rm pl}})^n]\,,
\label{vipo2}
\ee
where $0<\mu<1$ and $0<n<1$.
This potential behaves as $V(\phi) \simeq M^4[1-\mu(\phi/M_{\rm pl})^n]$
for $\phi \ll M_{\rm pl}$ and approaches $V(\phi)\to M^4 (1-\mu)$
in the limit $\phi \gg M_{\rm pl}$ (see Fig.~\ref{potential}).
For the potential (\ref{vipo2}) we obtain 
\be
\Gamma-1=\frac{1-nx-\mu (1-x)^n}{n \mu x (1-x)^n}\,,\qquad
x \equiv e^{-\phi/M_{\rm pl}}\,.
\ee
One can easily show that the r.h.s. is positive  
under the conditions $0<\mu<1, 0<n<1$,
so that $\Gamma>1$.
In both limits $\phi \to 0$ and $\phi \to \infty$
one has $\Gamma \to +\infty$. Since $\Gamma$ has a minimum
at a finite field value, the condition $\Gamma \gg 1$ is not necessarily 
satisfied today (unlike the potential (\ref{vipo1})).

Unless $\mu$ is very close to 1 the potential energy today 
is roughly of the order of $M^4$, i.e. $f(\phi_0) \approx 1$.
{}From Eq.~(\ref{con3}) it then follows that 
\be
n \mu (1-x_0)^{n-1}x_0 \simeq Q\,,
\label{nmure}
\ee
where $x_0 \equiv e^{-\phi_0/M_{\rm pl}}$.
If $\phi_0 \ll M_{\rm pl}$, we have that 
$\phi_0/M_{\rm pl} \simeq (n \mu)^{1/(1-n)}$.
{}From Eq.~(\ref{con2}) we obtain
\be
n \mu (1-x_B)^{n-1}x_B \simeq 10^{5}Q\,,
\label{nmure2}
\ee
where $x_B \equiv e^{-\phi_B/M_{\rm pl}}$.
Under the condition $\phi_B \ll M_{\rm pl}$ we have 
$\phi_B/M_{\rm pl} \simeq (10^{-5} n \mu/Q)^{1/(1-n)}$ from 
Eq.~(\ref{nmure2}).
Then the LLR bound (\ref{bou2}) corresponds to 
\be
n \cdot 10^{10-15n}<Q/\mu\,.
\label{lgcpo2}
\ee
When $\mu=0.5$ and $\mu=0.05$ with $Q=1$, the constraint 
(\ref{lgcpo2}) gives $n \gtrsim 0.63$ and 
$n \gtrsim 0.56$ respectively.

\subsection{Statefinder analysis}

The statefinder diagnostics introduced in Refs.~\cite{Sahni1,Sahni2} 
can be a useful tool to distinguish dark energy models from the 
$\Lambda$CDM model.
The statefinder parameters are defined by
\be
r=\frac{\dddot{a}}{aH^3}\,,\qquad
s=\frac{r-1}{3(q-1/2)}\,,
\label{sfinders}
\ee
where $q \equiv -\ddot{a}/(aH^2)$ is the deceleration parameter.
Defining $h \equiv H^2$, it follows that 
\be
q=-1-\frac{h'}{2h}\,,\qquad
r=1+ \frac{h''}{2h}+\frac{3h'}{2h}\,,
\ee
where a prime represents a derivative with respect to $N=\ln a$.

In the radiation dominated epoch we have $h \propto e^{-4N}$, 
which gives $(r, s) \simeq (3, 4/3)$.
During the matter era $r$ approaches $1$, whereas
$s$ blows up from positive to negative because of the divergence 
of the denominator in $s$ (i.e. $q=1/2$).
For the chameleon potentials (\ref{vipo1}) and (\ref{vipo2})
the solutions finally approach the de Sitter fixed point 
characterized by $(r, s)=(1,0)$. 
Around the de Sitter point the solutions 
evolve along the instantaneous minima characterized by 
$(x_1, x_2, x_3)=(\lambda/\sqrt{6}, \sqrt{1-\lambda^2/6}, 0)$
with $h \propto V$. Using Eq.~(\ref{lambdacou}) as well, one has 
$h'/h \simeq -\lambda^2$ and $h''/h \simeq (2\Gamma-1)\lambda^4$.
Then the statefinder diagnostics around the de Sitter point 
can be estimated as 
\bea
& & r \simeq 1+\left( \Gamma -\frac12 \right) \lambda^4
-\frac32 \lambda^2\,, \\
& & s \simeq -\frac{(2\Gamma-1)\lambda^4-3\lambda^2}
{3(3-\lambda^2)}\,.
\eea
%

\begin{figure}
\begin{centering}
\includegraphics[height=2.8in,width=2.8in]{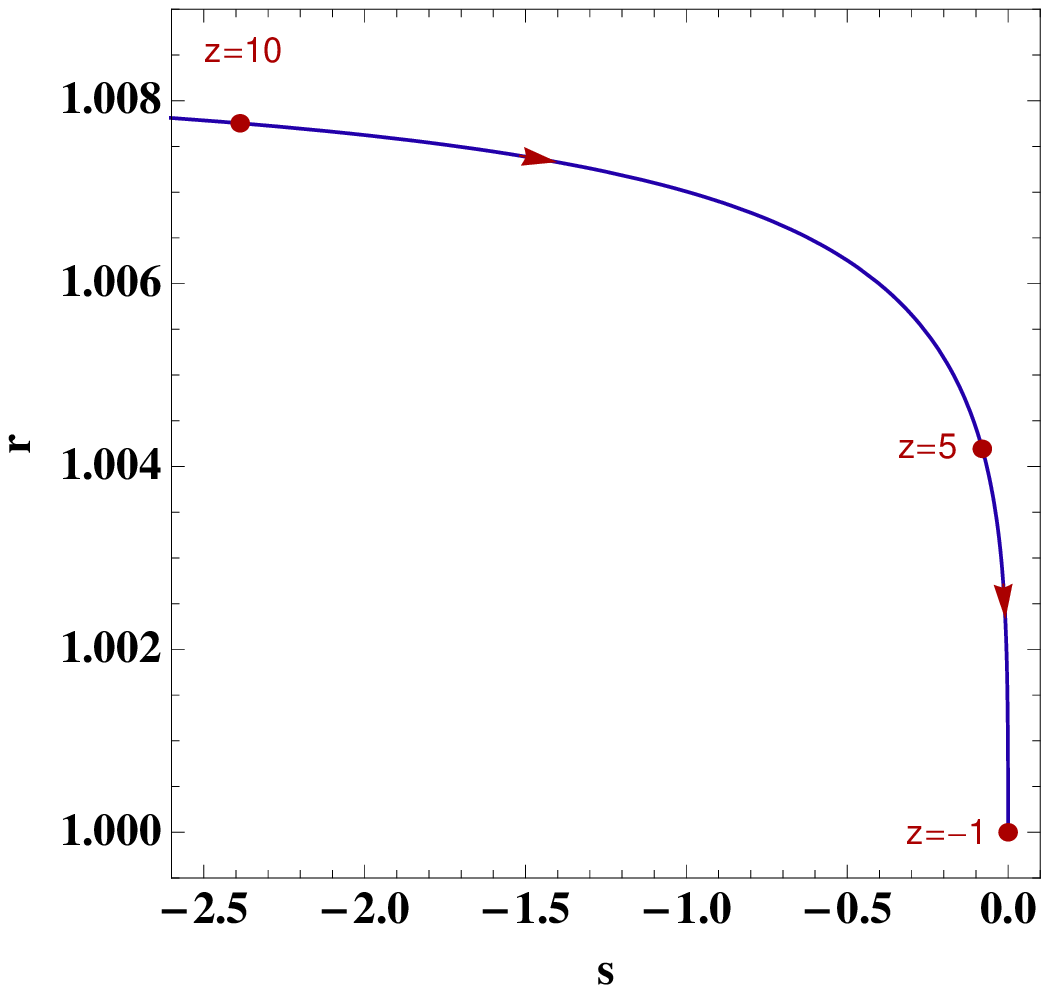}
\includegraphics[height=2.8in,width=2.8in]{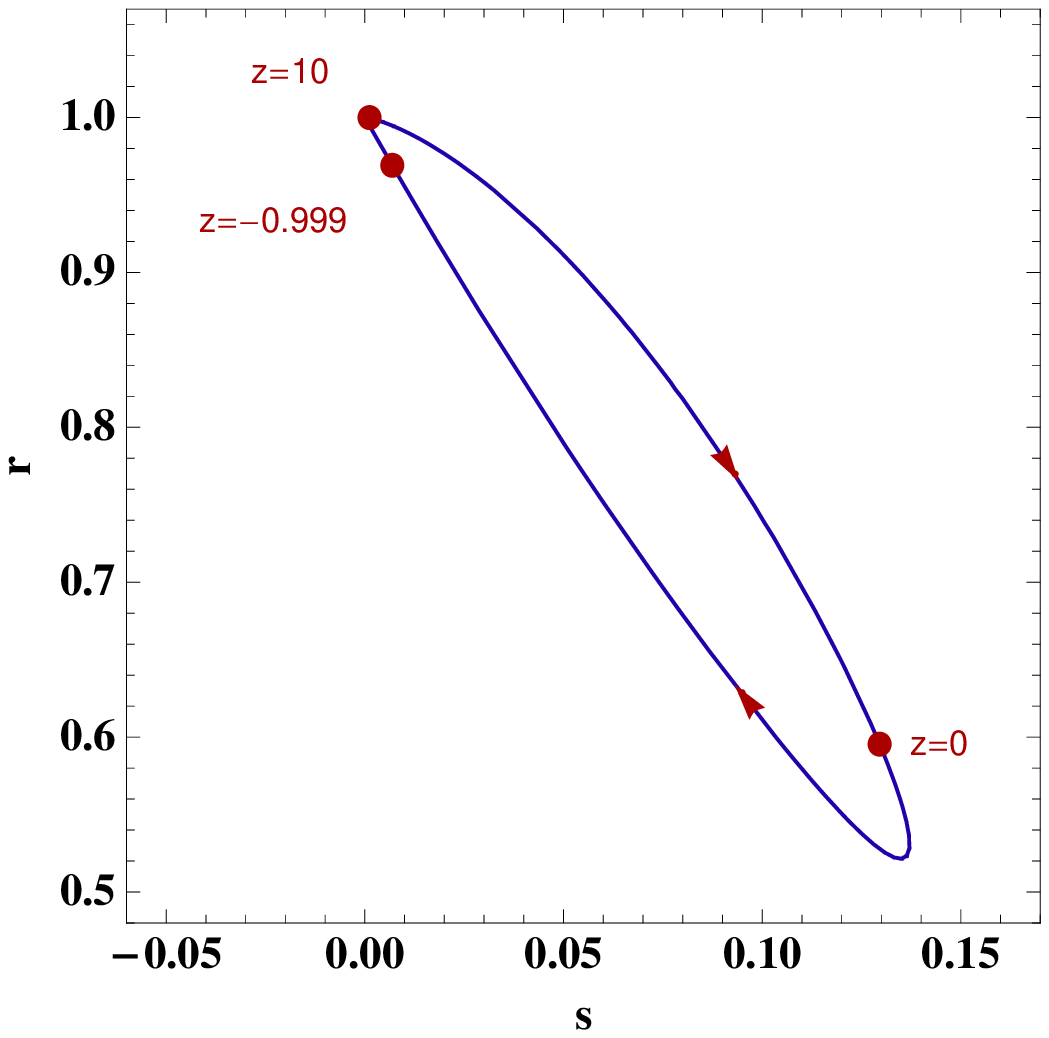}
\par\end{centering}
\caption{(Top): Evolution of the statefinders $r$ and $s$ 
for the potential (\ref{vipo1}) with $n=1$, $Q=1$ and $\mu=1$.
The solutions approach the de Sitter point at 
$(r, s)=(1, 0)$ from the region $r>1$ and $s<0$.
(Bottom): Evolution of statefinders for the potential 
(\ref{vipo2}) with $n=0.7$, $Q=1$ and $\mu=0.7$. In this case the solutions
approach the de Sitter point from the region $r<1$ and $s>0$.}
\label{statefinder}
\end{figure}

Since $\lambda$ finally approaches 0, Eq.~(\ref{lambdacou}) implies 
that $\Gamma \lambda^3 \to 0$ asymptotically.
For the potential in which $\Gamma \gg 1$ holds today 
it can happen that $\Gamma \lambda^4 \gg \lambda^2$, 
which gives $r \simeq 1+\Gamma \lambda^4>1$ and 
$s \simeq -2\Gamma \lambda^4/[3(3-\lambda^2)]<0$
around the present epoch.
In the upper panel of Fig.~\ref{statefinder} we plot the evolution 
of the variables $r$ and $s$ for the potential (\ref{vipo1})
in the redshift regime $-1<z \equiv 1/a-1<10$.
The statefinders evolve toward the de Sitter point characterized 
by $(r,s)=(1,0)$ from the regime $r>1$ and $s<0$.
This behavior is different from quintessence with the 
power-law potential $V(\phi)=M^{4+n}\phi^{-n}$ ($n>0$)
in which the statefinders are confined in the
region $r<1$ and $s>0$ \cite{Sahni1,Sahni2}.

The potential (\ref{vipo2}) allows the possibility that $\Gamma$
is not much larger than 1 even at the present epoch.
In the regime $\lambda^2 \ll 1$ we then have 
$r \simeq 1-3\lambda^2/2<1$ and $s \simeq \lambda^2/3>0$.
In fact we have numerically confirmed that the solutions enter
this regime by today (see the lower panel of Fig.~\ref{statefinder}). 
Finally they approach the de Sitter point from 
the regime $r<1$ and $s>0$.
Hence one can distinguish between chameleon potentials from 
the evolution of statefinders.

\section{Local gravity constraints on chameleon potentials}
\label{lgcsec}

In this section we discuss a number of local gravity constraints on 
the chameleon potentials (\ref{vipo1}) and (\ref{vipo2}) in details.
Together with the LLR bound (\ref{bou2}) we use the constraint
coming from 2006 E\"{o}t-Wash experiments \cite{Kapner:2006si} 
as well as the WMAP bound on the variation of 
the field-dependent mass. 

\subsection{The WMAP constraint on the variation of the particle mass}

Due to the conformal coupling of the field $\phi$ to matter, 
any particle will acquire a $\phi$-dependent mass:
\be
m(\phi) = m_0 e^{Q\phi/M_{\rm pl}}\,,
\ee
where $m_0$ is a constant.

The WMAP data constrain any variation in $m(\phi)$, between
now and the epoch of recombination to be $\lesssim 5\%$ at $2\sigma$ 
($\lesssim 23\%$ at $4\sigma$) \cite{wmap_constraints}. 
We then require that 
\be
\left|\frac{\Delta m(\phi)}{m}\right| =
\left| e^{\frac{Q(\phi_0-\phi_{\rm rec})}
{M_{\rm pl}}}-1 \right| \lesssim 0.05\,,
\ee
where $\phi_{\rm rec}$ is the field value at the recombination epoch.
If we assume that the chameleon follows the minimum since recombination then
$\phi_0\gg \phi_{\rm rec}$, and the field in the cosmological background today
must satisfy
\be
Q\phi_0/M_{\rm pl} \lesssim 0.05\,.
\label{WMAP}
\ee
This provides a constraint on the coupling $Q$ and 
the model parameters of chameleon potentials. 

Note that the WMAP constraint is not a local gravity constraint.  
Nevertheless, it provides strong constraints on the potential (\ref{vipo2}) 
with natural parameters and is therefore considered in this section.

\subsection{Constraints from the 2006 E\"{o}t-Wash experiment}

The 2006 E\"{o}t-Wash experiment \cite{Kapner:2006si} searched for deviations
from the $1/r^2$ force law of gravity. The experiment used two parallel plates,
the detector and attractor, which are separated by a (smallest) distance
$d=55\mu$m. The plates have holes of different sizes bored into them, and the
attractor is rotating with an angular velocity $\omega$. The rotation of the
attractor gives rise to a torque on the detector, and the setup of the
experiment is such that this torque vanishes for any force that falls off as
$1/r^2$. In between the plates there is a $d_s = 10\,\mu$m BeCu-sheet, which
is for shielding the detector from electrostatic forces.

The chameleon force between two parallel plates, see e.g.
Ref.~\cite{Brax:2007vm}, usually falls off faster than $1/r^2$, implying a
strong signature on the experiment. However, if the matter-coupling is strong
enough, the electrostatic shield will itself develop a thin-shell. When this
happens, the effect of this shield is not only to shield electrostatic forces,
but also to shield the chameleon force on the detector. This suppression is
approximately given by a factor $\exp(-m_s d_s)$, where $m_s$
is the mass inside the electrostatic shield. 
Hence the experiment cannot detect strongly coupled chameleons.

The behavior of chameleons in the E\"{o}t-Wash experiment have been explained in
Refs.~\cite{Mota:2006fz,Brax:2008hh,Brax:2010kv}. 
We calculate the E\"{o}t-Wash constraints on our models numerically based 
on the prescription presented in Ref.~\cite{Brax:2008hh}.

\subsection{Combined local gravity constraints}
\subsubsection{Potential $V(\phi)=M^4 \exp[\mu (M/\phi)^n]$}
%

\begin{figure}
\begin{centering}
\includegraphics[height=2.9in,width=3.0in]{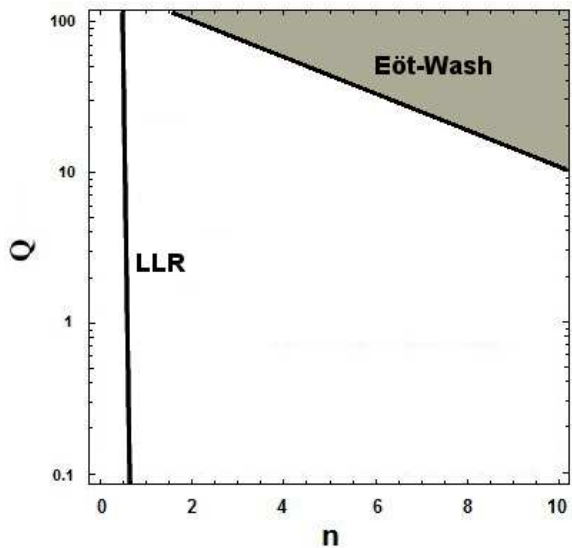}
\par\end{centering}
\caption{The combined local gravity constraints 
on the potential (\ref{vipo1}) with $\mu=1$
in the $(n,Q)$ plane. The shaded region corresponds 
to the allowed parameter space.
The natural values of $Q$ and $n$ of the order 
of unity are excluded.}
\label{fig3}
\end{figure}

Let us first consider the inverse-power law potential (\ref{vipo1})
with $n>0$.
{}From Eq.~(\ref{extremum}) the field value $\phi_m$ at the 
minimum of the effective potential $V_{\rm eff}$ satisfies
\be
\left( \frac{M}{\phi_m} \right)^{n+1}=\frac{Q}{\mu n}
\frac{M}{M_{\rm pl}} \frac{\rho_m e^{Q\phi_m/M_{\rm pl}}}
{V(\phi_m)}\,.
\ee
In this model the field is in the regime $M \ll \phi_m \ll M_{\rm pl}$
for the density $\rho_m$ we are interested in.
Since $V(\phi_m)$ can be approximated as $V(\phi_m) \simeq M^4$, 
it follows that 
\be
\frac{\phi_m}{M} \simeq \left( \frac{Q}{\mu n}
\frac{M}{M_{\rm pl}} \frac{\rho_m}{M^4} \right)^{-1/(n+1)}\,.
\ee

Using the LLR bound (\ref{bou2}) with the homogeneous density 
$\rho_m \simeq 10^{-24}$\,g/cm$^3$ in our galaxy, we obtain 
the constraint 
\be
n \cdot 10^{10-15n}<Q/\mu\,.
\label{lgcpo1}
\ee

The WMAP bound (\ref{WMAP}) gives
\be
Q< \frac{M_{\rm pl}}{M} \left( \frac{0.05^{n+1}}
{\mu n} \frac{\rho_m^{(0)}}{M^4} \right)^{1/n}\,,
\label{WMAP1}
\ee
where $\rho_m^{(0)}$ is the matter density today, with 
$\rho_m^{(0)}/M^4 \approx \Omega_{m}^{*(0)}/\Omega_{\phi}^{(0)}
\approx 1/3$.
Since $M_{\rm pl}/M \approx 10^{30}$, this condition 
is well satisfied for $Q, n, \mu$ of the order of unity.

For the E\"{o}t-Wash experiment, the chameleon torque on the detector 
was found numerically to be larger than the experimental bound 
when $Q, n, \mu$ are of the order of unity.
Providing the electrostatic shield with a thin-shell, 
we require that $n\gg 1$, $Q\gg 1$ or $\mu\ll 1$ 
to satisfy the experimental bound.

In Fig.~\ref{fig3} we plot the region constrained by 
the bounds (\ref{lgcpo1}), (\ref{WMAP1}), 
and the E\"{o}t-Wash experiment for $\mu=1$.
This shows that only the large coupling region with $Q \gg 1$
can be allowed for $n$ of the order of unity.
A viable model can also be constructed by taking values of $\mu$ 
much smaller than 1.
Note that the WMAP bound (\ref{WMAP1}) is satisfied 
for the parameter regime shown in Fig.~\ref{fig3}.

\subsubsection{Potential $V(\phi)=M^4 [ 1-\mu (1-e^{-\phi/M_{\rm pl}})^n]$}

Let us proceed to another potential (\ref{vipo2}) with $0<n<1$.
In the regions of high density where local gravity experiments are 
carried out, we have $\phi \ll M_{\rm pl}$ and hence
$V(\phi) \simeq M^4 [1-\mu (\phi/M_{\rm pl})^n]$.
In this regime the effective potential $V_{\rm eff}$ has 
a minimum at 
\be
\label{phimin_vipo2}
\phi_m=\left( \frac{Q}{\mu n} \frac{\rho_m}{M^4} 
\right)^{1/(n-1)}M_{\rm pl}\,.
\ee

Recall that the LLR bound was already derived in Eq.~(\ref{lgcpo2}), 
which is the same as the constraint (\ref{lgcpo1}) 
of the previous potential.

Assuming that the chameleon is at the minimum of its effective potential 
in the cosmological background today, 
the WMAP bound (\ref{WMAP}) translates into 
\be
Q \gtrsim 0.05 (60n \mu)^{1/n}\,,
\label{WMAP2}
\ee
where we have used $\rho_{m}^{(0)}/M^4 \approx 1/3$ 
in Eq.~(\ref{phimin_vipo2}). However, a full numerical simulation of the
background evolution shows that this is not always the case. For a large range
of parameters the chameleon has started to lag behind the minimum,
which again leads to a weaker constraint.

The E\"{o}t-Wash experiment provides the strongest constraints 
when $Q$ is of the order of unity for the potentials 
(\ref{fphi}) and (\ref{vipo1}).
This is not the case for the potential (\ref{vipo2}), 
because the electrostatic shield used 
in the experiment develops a thin-shell.

The mass inside the electrostatic shield is given by
\be
m_s^2 \simeq n(1-n)\mu\left(\frac{Q \rho_s}{\mu n
M^4}\right)^{\frac{2-n}{1-n}}\frac{M^4}{M_{\rm pl}^2}\,.
\ee
Using $\rho_s \simeq 10$\,{\rm g/cm}$^3$ and 
$d_s \simeq 10\,\mu$m we have
\be
m_s d_s\simeq \sqrt{n(1-n)\mu}\left(\frac{Q}{\mu
n}\right)^{\frac{2-n}{1-n}}10^{\frac{6-35n}{1-n}}\,.
\ee
Taking $Q$ and $\mu$ to be of the order of unity,
we find $m_s d_s \gg 1$ as long as $n \gtrsim 0.2$.
The suppression of the chameleon torque due to the presence
of the electrostatic shield makes the chameleon 
invisible in the experiment.

\begin{figure}
\begin{centering}
\includegraphics[height=2.9in,width=3.0in]{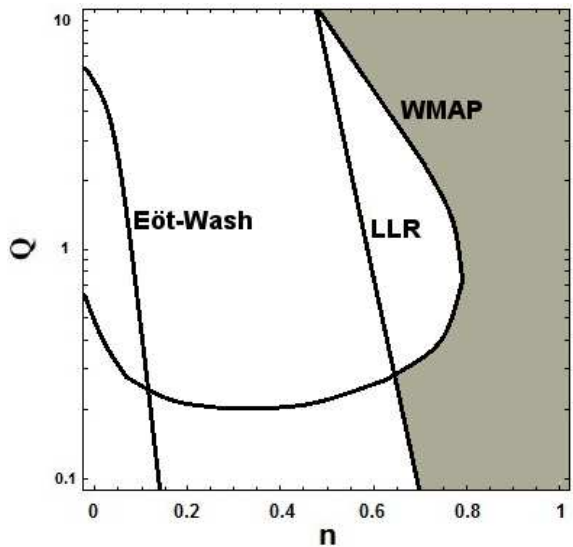}
\par\end{centering}
\caption{The combined local gravity constraints 
on the potential (\ref{vipo2}) with $\mu=0.5$
in the $(n,Q)$ plane. In this case the allowed parameter 
space (shaded region in the figure) is determined by the 
WMAP constraint and the LLR constraint.}
\label{fig4}
\end{figure}

\begin{figure}
\begin{centering}
\includegraphics[height=2.9in,width=3.0in]{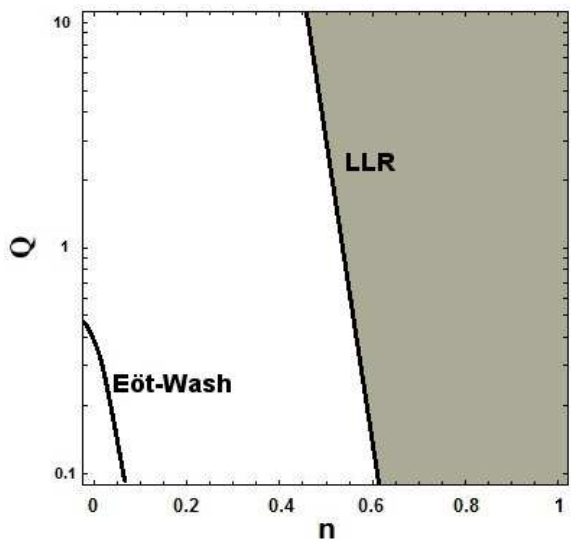}
\par\end{centering}
\caption{The combined local gravity constraints 
on the potential (\ref{vipo2}) with $\mu=0.05$.
The allowed parameter space is determined by 
the LLR bound. The WMAP constraint 
is satisfied for the whole parameter space 
in the figure. }
\label{fig5}
\end{figure}

In Figs.~\ref{fig4} and \ref{fig5} we plot the allowed regions
constrained by the bounds (\ref{lgcpo2}), (\ref{WMAP2}), and 
the E\"{o}t-Wash experiments, for $\mu=0.5$ and $\mu=0.05$, 
respectively. When $\mu=0.5$ the WMAP constraint gives the 
tightest bound for $Q \gtrsim 0.2$, and the parameter space 
$Q \lesssim 1$ is viable for $n \gtrsim 0.7$. 
If we decrease the values of $\mu$ down to 0.05, then the 
WMAP bound is well satisfied for the parameter space shown 
in Fig.~\ref{fig5}. Instead, the LLR experiment provides the 
tightest bound in such cases.
When $\mu=0.05$, the region with $0.1 \lesssim Q \lesssim 10$
and $n \gtrsim 0.6$ can be allowed.
The coupling $Q$ as well as the parameter $n$ are
not severely constrained for the potential (\ref{vipo2}).

\section{Linear growth of matter perturbations}

We now turn our attention to cosmological perturbations 
in chameleon cosmology. It is well-known that matter perturbations 
allow to discriminate between dark energy models where 
the gravitational interaction is modified on cosmic scales.
We first review the general formalism and derive the equation for
linear matter perturbations. We also introduce important 
quantities like the critical scale $\lambda_c$ below which 
modifications of gravity are felt and the growth index 
$\gamma(z,k)$, a powerful discriminative quantity for the 
study of the modified evolution of matter perturbations as was 
explained in the Introduction.  

As we have seen in Sec.~\ref{lgcsec}, local gravity constraints impose very
strong boundaries on the potential (\ref{vipo1}), 
forcing its parameters to take unnatural values.
Moreover, for viable choices of parameters, we have verified that the linear 
perturbations behave in a manner similar to the $\Lambda$CDM model,
as $\lambda_c$ is much smaller than the cosmic scales we are interested 
in. On the other hand we have shown that the model parameters of 
the potential (\ref{vipo2}) is not severely constrained.
In Sec.~\ref{nonuni} we will show that the potential (\ref{vipo2})
gives rise to some very interesting observational signatures. 

\subsection{General formalism for cosmological perturbations}  
\label{genefor}

We consider scalar metric perturbations $\alpha$, $B$, 
$\psi$, and $\gamma$ around a flat FLRW background.
The line-element describing such a perturbed Universe is 
given by \cite{Bardeen}
\bea
{\rd} s^2&=&-(1+2\alpha){\rm d}t^2-2 a B_{,i}{\rm d} t {\rm d} x^i
\nonumber \\ & &
+a(t)^2\left[ (1+2\psi)\delta_{ij}+2\gamma_{,i;j} \right]
{\rm d} x^i {\rm d} x^j\,.
\eea
We decompose the field $\phi$ into the background and inhomogeneous
parts: $\phi (t, {\bm x})=\tilde{\phi}(t)+\delta \phi (t, {\bm x})$.
The energy-momentum tensors $T_{\mu \nu}^{(m)}$ of non-relativistic 
matter can be decomposed as
\be
{T^{0}_{0}}^{(m)}=-(\rho_m^*+\delta \rho_m^*)\,,\qquad
{T^0_{i}}^{(m)}=-\rho_m^* v_{,i}\,,
\ee
where $v$ is the peculiar velocity potential of non-relativistic 
matter. In the following, when we express background quantities, 
we drop the tilde for simplicity.

Let us consider the evolution of matter perturbations, 
$\delta_m \equiv \delta \rho_m^*/\rho_m^*$ 
in the comoving gauge ($v=0$). 
The quantity $\delta_m$ corresponds to the gauge-invariant 
quantity introduced in Refs.~\cite{Bardeen,S98,BEPS00} 
when expressed in the comoving gauge.  
In the Fourier space the first-order perturbation equations are 
given by \cite{Hwang:1991aj,Amendolaper}
\begin{widetext}
\bea
& & \alpha=-Q \delta \phi/M_{\rm pl}\,,
\label{per1} \\
& & \dot{\delta \rho^*_m} + 3 H \delta\rho^*_m - \rho^*_m \left( \kappa - 3H
\alpha
\right)- Q (\rho^*_m \dot{\delta\phi} + \delta\rho^*_m \dot{\phi})/M_{\rm
pl}=0\,,
\label{per2} \\
& & \ddot{\delta \phi}+3H \dot{\delta \phi}+\left( V_{,\phi
\phi}+\frac{k^2}{a^2}
\right) \delta \phi+2\alpha V_{,\phi}-\dot{\phi} (\dot{\alpha}-3H \alpha+\kappa)
+Q (2\alpha \rho_m^*+\delta \rho_m^*)/M_{\rm pl}=0\,,
\label{per3} \\
& &
\dot{\kappa} + 2H\kappa + 3\dot{H} \alpha-\frac{k^2}{a^2}\alpha 
-\frac{1}{2M_{\rm pl}^2} \left( \delta\rho^*_m -4\alpha\dot{\phi}^2 
+4\dot{\phi}\dot{\delta \phi} - 2V_{,\phi}\delta\phi\right)=0\,,
\label{per4} 
\eea 
\end{widetext}
where $k$ is a comoving wavenumber and
$\kappa \equiv 3(H \alpha -\dot{\psi}) +
(k^2/a) (B+a \dot{\gamma})$.
{}From Eq.~(\ref{per2}) it follows that
\be
\kappa=\dot{\delta}_m-Q (\dot{\delta \phi}+
3H \delta \phi)/M_{\rm pl}\,,
\label{kappa}
\ee
where we have used Eq.~(\ref{per1}).
Plugging Eq.~(\ref{kappa}) into Eqs.~(\ref{per3}) and (\ref{per4}), 
we obtain 
\begin{widetext}
\bea
& & \ddot{\delta \phi}+\left( 3H+2Q\frac{\dot{\phi}}{M_{\rm pl}}
\right) \dot{\delta \phi}+\left( m_{\phi}^2+\frac{k^2}{a^2}
-2Q \frac{V_{,\phi}}{M_{\rm pl}}-2Q^2 \frac{\rho_m^*}{M_{\rm pl}^2}
\right) \delta \phi+\frac{Q}{M_{\rm pl}} \rho_m^* \delta_m
-\dot{\phi}\,\dot{\delta}_m=0\,,
\label{delphi}
\\
& &\ddot{\delta}_m+\left( 2H-Q\frac{\dot{\phi}}{M_{\rm pl}} \right)
\dot{\delta}_m-\frac{\rho_m^* \delta_m}{2M_{\rm pl}^2}(1-2Q^2)
+\biggl[ \frac{V_{,\phi}}{M_{\rm pl}^2}+\frac{Q}{M_{\rm pl}}
\biggl(m_{\phi}^2+\frac{2k^2}{a^2}-6H^2-6\dot{H}-
\frac{2\dot{\phi}^2}{M_{\rm pl}^2} \nonumber \\
& &-2Q \frac{V_{,\phi}}{M_{\rm pl}}
-2Q^2 \frac{\rho_m^*}{M_{\rm pl}^2}\biggr) \biggr] \delta \phi
+\frac{1}{M_{\rm pl}} \left( 2Q^2 \frac{\dot{\phi}}{M_{\rm pl}}
-\frac{2\dot{\phi}}{M_{\rm pl}}-2QH \right) \dot{\delta \phi}=0\,,
\label{deldeltam}
\eea 
\end{widetext}
where $m_\phi^2=V_{,\phi\phi}$ is the 
mass squared of the chameleon field.

As long as the field $\phi$ evolves slowly (``adiabatically'') 
along the instantaneous minima of 
the effective potential $V_{\rm eff}$, one can employ the quasi-static
approximation on sub-horizon 
scales ($k \gg aH$) \cite{S98,BEPS00,quasi}.
This corresponds to the approximation under which the dominant terms 
in Eqs.~(\ref{delphi}) and (\ref{deldeltam}) are those including 
$k^2/a^2$, $m_{\phi}$, and $\delta_m$, i.e.
\bea
& & \left( m_{\phi}^2+\frac{k^2}{a^2} \right) \delta \phi \simeq 
-\frac{Q}{M_{\rm pl}} \rho_m^* \delta_m\,,\\
& & \ddot{\delta}_m + 2H\dot{\delta}_m-\frac{\rho_m^* \delta_m}
{2M_{\rm pl}^2} (1-2Q^2) \nonumber \\
& &+\frac{Q}{M_{\rm pl}} 
\left( m_{\phi}^2+\frac{2k^2}{a^2} \right) \delta \phi \simeq 0\,,
\eea
where we have also used the approximation $\dot{\phi}/M_{\rm pl} \ll H$.
Combining these equations, it follows that 
\be
\label{delta}
\ddot{\delta}_m + 2H\dot{\delta}_m 
-4\pi G_{{\rm eff}}\rho_m^* \delta_m 
\simeq 0\,,
\ee
where the effective gravitational coupling is given by 
\be
\label{del}
G_{{\rm eff}}=G \left( 1+2Q^2 \frac{k^2/a^2}
{ m_\phi^2 +k^2/a^2} \right)\,.
\ee
An analogous modified equation was found in Refs.~\cite{TUMTY,BEPS00,Song},
the crucial point being to elucidate the physical significance of $G_{\rm eff}$. 
We can understand the physical content of the modification of
gravity by looking at the corresponding gravitational 
potential in real space. 
The gravitational potential (per unit mass) is of the type
$V(r) =-(G /r)\left( 1 + 
2Q^2\,e^{- m_{\phi} r} \right)$ \cite{darkbook}.

When we solve the full system of perturbations (\ref{deldeltam}), 
we can, for some of our models, get a small discrepancy compared to (\ref{delta}). 
This can result in a non-negligible difference of up to around $5\%$ in the numerical calculation of the growth rate of matter perturbations.
This arises mainly because the field $\phi$ does not 
move exactly along the minimum of the effective potential but is instead lagging 
a little behind it. 

We see that in chameleon models $G_{\rm eff}$ is a scale-dependent 
as well as a time-dependent quantity.
Clearly the scale-dependent driving force in Eq.~(\ref{del}) 
induces in turn a scale dependence in the growth of matter 
perturbations with two asymptotic regimes, i.e.
\begin{align}
G_{\rm eff} &= G ( 1 + 2Q^2 ) & k/a& \gg m_{\phi}~~{\rm or}~~
\lambda \ll \lambda_c ~,\lb{as1}\\ 
&= G & k/a &\ll m_{\phi}~~
{\rm or}~~\lambda \gg \lambda_c~,\lb{as2}
\end{align}
where we have introduced the physical wavelength $\lambda=(2\pi/k)a$.
We have in particular $\lambda_0= 2\pi/k$ today ($a=1$). 
The characteristic (physical) scale $\lambda_c$ is defined by 
\be
\lambda_c=2\pi/m_{\phi}\,. 
\ee
On scales $\lambda \gg \lambda_c$ matter perturbations do not feel 
the fifth force during their growth. 
On the contrary, on scales much smaller than $\lambda_c$ 
they do feel its presence. 
During the matter dominance ($\Omega_m^* \simeq 1$) the solutions to 
Eq.~(\ref{delta}) are given by 
\begin{align}
\delta_m  &\propto a^{[ \sqrt{1+24(1+2Q^2)}-1]/4} 
                   & \lambda & \ll \lambda_c\,, \lb{delas1}\\
\delta_m  &\propto a & \lambda & \gg \lambda_c\,.\lb{delas2} 
\end{align}
Hence, in the regime $\lambda \ll \lambda_c$, the growth rate
gets larger than that in standard General Relativity.

As mentioned earlier, a powerful way to describe the growth of 
perturbations is by introducing the function $\gamma(k,z)$ defined 
as follows 
\be
f=\Omega^*_m(z)^{\gamma(k,z)}~,
\ee 
where 
\be
f=\frac{{\rm d} \ln \delta_m}
{{\rm d} \ln a}\,.
\ee
We remind the definition $\delta_m \equiv \delta \rho_m^*/\rho_m^*$.
The quantity $\gamma$ can be time-dependent and also
scale-dependent.
It is known that a large class of dark energy models 
inside General Relativity yields a quasi-constant $\gamma$ 
with values close to that of the $\Lambda$CDM model, 
$\gamma\approx 0.55$ \cite{Linder,PG07}.
Therefore any significant deviation from this behavior would 
give rise to a characteristic signature for our chameleon models. 
Since $0<\Omega^*_m(z)<1$, smaller $\gamma$ implies 
a larger growth rate of matter perturbations.

As we have seen before, the chameleon mechanism is devised 
so that in high-density environments the mass of 
the scalar field is large relative 
to its value in low-density ones. During the cosmological evolution, 
the mass of the field will follow this behavior, which 
means that $\lambda_c$ will move from small to large values. 
In other words, $G_{\rm eff}$ will evolve from the regime (\ref{as2}) 
to the regime (\ref{as1}). 
This transition is scale-dependent, which is an important feature of 
the growth of matter perturbations in chameleon (and $f(R)$) models. 

We consider the evolution of matter perturbations for the wavenumbers 
\begin{equation}
0.01~h\,{\rm Mpc}^{-1} \lesssim k \lesssim 
0.2~h\,{\rm Mpc}^{-1}\,,
\label{krange}
\end{equation}
where $h$ describes the uncertainty of the Hubble parameter $H_0$ today,
i.e. $H_0=100\,h$\,km\,sec$^{-1}$\,Mpc$^{-1}$.
The scales (\ref{krange}) range from the upper limit of observable scales in
the linear regime of perturbations to the mildly non-linear regime
(in which the linear approximation is still reasonable).

Depending on the value of $\lambda_c$ today (denoted as
$\lambda_{c,0}$) and on its recent evolution, three
possibilities can actually arise: (i) The model is hardly distinguishable from 
$\Lambda$CDM; (ii) The model is distinguishable from $\Lambda$CDM but shows no 
dispersion i.e. no scale-dependence. In this case low values of $\gamma_0$ 
will also yield large slopes 
$\gamma'_0 \equiv ({\rm d}\gamma/{\rm d}z)(z=0)$, much larger 
than in $\Lambda$CDM; and finally (iii) The model is distinguishable from 
$\Lambda$CDM and shows some dispersion altogether. These three cases can be 
characterized using the quantity $\gamma_0\equiv \gamma(z=0)$. 
This classification is analogous to what was done for some viable 
$f(R)$ models \cite{Tsujikawa:2009ku} and it can be defined as follows:
\begin{itemize}
\item \rm{(i)} The region in parameter space for which $\gamma_0>0.53$ 
for all the scales described by (\ref{krange}).
In this region $\gamma'_0$ is small and $\gamma$ is 
nearly constant.  
\item \rm{(ii)} The region where $\gamma_0$ is degenerate, i.e. assumes 
the same value for all the scales, with a value smaller than $0.5$. In this 
region $\gamma'_0$ is large and there is a significant variation of 
$\gamma$.  
\item \rm{(iii)} The region where $\gamma_0$ shows some dispersion, i.e. 
a scale-dependence. For low $\gamma_0$, $\gamma'_0$ is large and we have 
significant changes of $\gamma$. 
\end{itemize}
Models in the regions (ii) and (iii) can be clearly discriminated from $\Lambda$CDM. 
Some examples are shown in Figs.~\ref{growth1} and \ref{growth2}.
We investigate below in more details the appearance of these 
characteristic signatures. 

\subsection{Observational signatures in the growth of matter perturbations}
\label{nonuni}

The main question when looking at linear perturbations in chameleon 
models is the order of magnitude of the scale $\lambda_c$.
If the chameleon mass $m_{\phi}$ is large such that $\lambda_c$ 
is less than the order of the galactic size, matter perturbations 
on the scales relevant to large scale structures
do not feel the chameleon's presence. 
This is actually the case for the inverse power 
exponential potential (\ref{vipo1}) with model 
parameters bounded by observational and 
local gravity constraints.
Therefore, we will concentrate the analysis of the perturbations 
on the potential (\ref{vipo2}).

In the regime $\phi/M_{\rm pl} \ll 1$ one can employ the 
approximation $V(\phi) \simeq M^4[1-\mu(\phi/M_{\rm pl})^n]$. 
In fact this approximation is valid for most of the cosmological 
evolution by today. Then we obtain the field mass at the 
minimum of the effective potential $V_{\rm eff}(\phi)$:
\be
m_{\phi} \simeq \frac{2\pi}{\lambda_{c,0}}
e^{-\frac{3(2-n)}{2(1-n)}N}\,,
\label{mphi}
\ee
where $\lambda_{c,0}$ is the critical length today, given by 
\be
\lambda_{c,0}=\frac{2\pi}{\sqrt{\mu n(1-n)}} \frac{M_{\rm pl}}
{M^2} \left( \frac{Q}{\mu n} \frac{\rho_m^{(0)}}{M^4} 
\right)^{-\frac{2-n}{2(1-n)}}\,.
\ee
Using the relation $\rho_{\rm DE}^{(0)}=
3M_{\rm pl}^2 H_0^2 \Omega_{\rm DE}^{(0)} \simeq M^4$, 
we find that $\lambda_{c,0}$ is at most of the order of $H_0^{-1}$.
If $Q=1$, $n=0.6$, $\mu=0.05$, and $\Omega_{\rm DE}^{(0)}=0.72$, 
for example, $\lambda_{c,0} \approx 0.4 H_0^{-1}$.
For the modes deep inside the Hubble radius today ($\lambda_0 \ll H_0^{-1}$)
the perturbations are affected by the modification of gravity.

Plugging Eq.~(\ref{mphi}) into Eq.~(\ref{del}), the effective gravitational 
coupling can be expressed in the form
\be
G_{\rm eff}= G \left [1+\frac{2Q^2}
{1+e^{-A(N-N_c)}}\right]\,,
\ee
where 
\bea
A &=& \frac{4-n}{1-n}\,,\label{param1}\\
N_c &=& \frac{2(1-n)}{4-n}\ln
\left(\frac{\lambda_0}{\lambda_{c,0}}\right)\,.
\label{param2}
\eea
Here $A$ and $N_c$ are the parameters describing
respectively the steepness of transition from the inferior to the superior asymptote 
and the position of the half-amplitude value of the 
function [i.e. $G_{\rm eff}=G (1+Q^2)$]. 

\begin{figure}
\includegraphics[scale=.7]{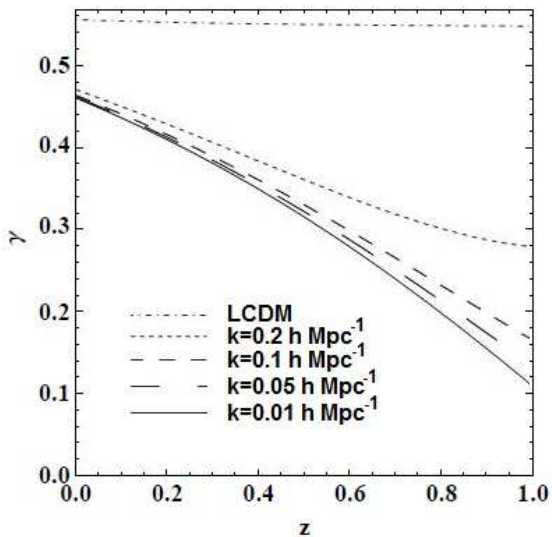}
\caption{\label{growth1} 
The evolution of the growth index $\gamma(z)$ versus the redshift $z$ 
in the model (\ref{vipo2}) with $n=0.8$, $\mu=0.5$ and $Q=1/\sqrt{6}$ 
for four different values of $k$. 
For these parameters, the model passes local gravity constraints. 
In this case the field is lagging behind the minimum, 
which means that the approximated equation (\ref{delta}) provides the results 
slightly different from this plot (by around 4\% in the value of $\gamma_0$).}
\end{figure}

\begin{figure}
\includegraphics[scale=.7]{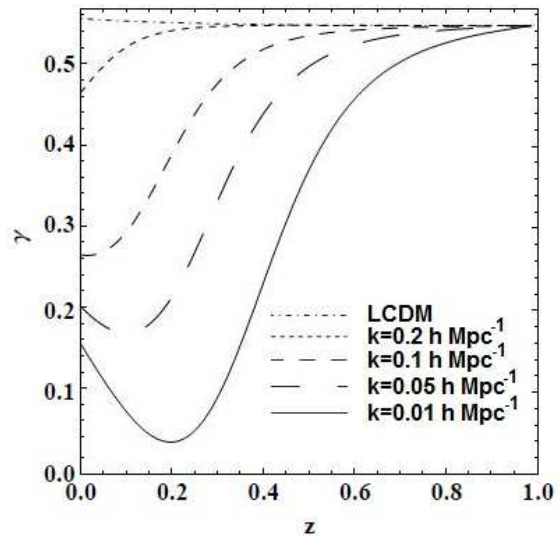}
\caption{\label{growth2} 
The evolution of $\gamma$ versus the redshift $z$ in the model (\ref{vipo2}) 
with $n=0.7$, $\mu=0.05$ and $Q=1$ for four different values of $k$.
In this case the model passes local gravity constraints, too. Contrary
to Fig.~\ref{growth1} we have a significant dispersion of $\gamma$, 
depending on the wavenumbers $k$.}
\end{figure}

The first interesting point to remark is that the steepness of transition
depends exclusively on the value of $n$, with a step function as a limit when
$n\rightarrow 1$ and a slower transition for $n\rightarrow 0$. Nonetheless, it
must be remembered that this is so in the variable $N$. When converting back to
the redshift $z$, the logarithm scale introduces a distortion, which means that
the earlier a scale starts its transition, the slower this transition will be.
In this sense, there is a scale-dependence to the steepness of the transition,
even if it does not appear explicitly in Eq.~(\ref{param1}).

If we want to get a better feel for the epoch of the  transition, 
we can define the redshift $z_c={\rm e}^{-N_c}-1$.
{}From Eq.~(\ref{param2}) it follows that 
\be
z_c=\left (\frac{\lambda_{c,0}}{\lambda_0}
\right)^{\frac{2(1-n)}{4-n}}-1\,.
\label{transrshift}
\ee
As expected, the transition redshift is scale-dependent, with higher values for
the smaller scales. Moreover, $z_c\rightarrow 0$ for all scales when
$n\rightarrow 1$. Since this is also the limit at which the transition becomes a
step function, we have thus an indication that a transition that happens very
close to the present will necessarily be very steep. Another remark is that 
the transition happens in the past ($z_c>0$) for the scales $\lambda_0$
smaller than $\lambda_{c,0}$.
Thus $\lambda_{c,0}$ is a good indication of the scale around which 
there will be a dispersion, as it marks the scale that will be exactly 
in the middle of the transition today.

\begin{figure}
\includegraphics[scale=.7]{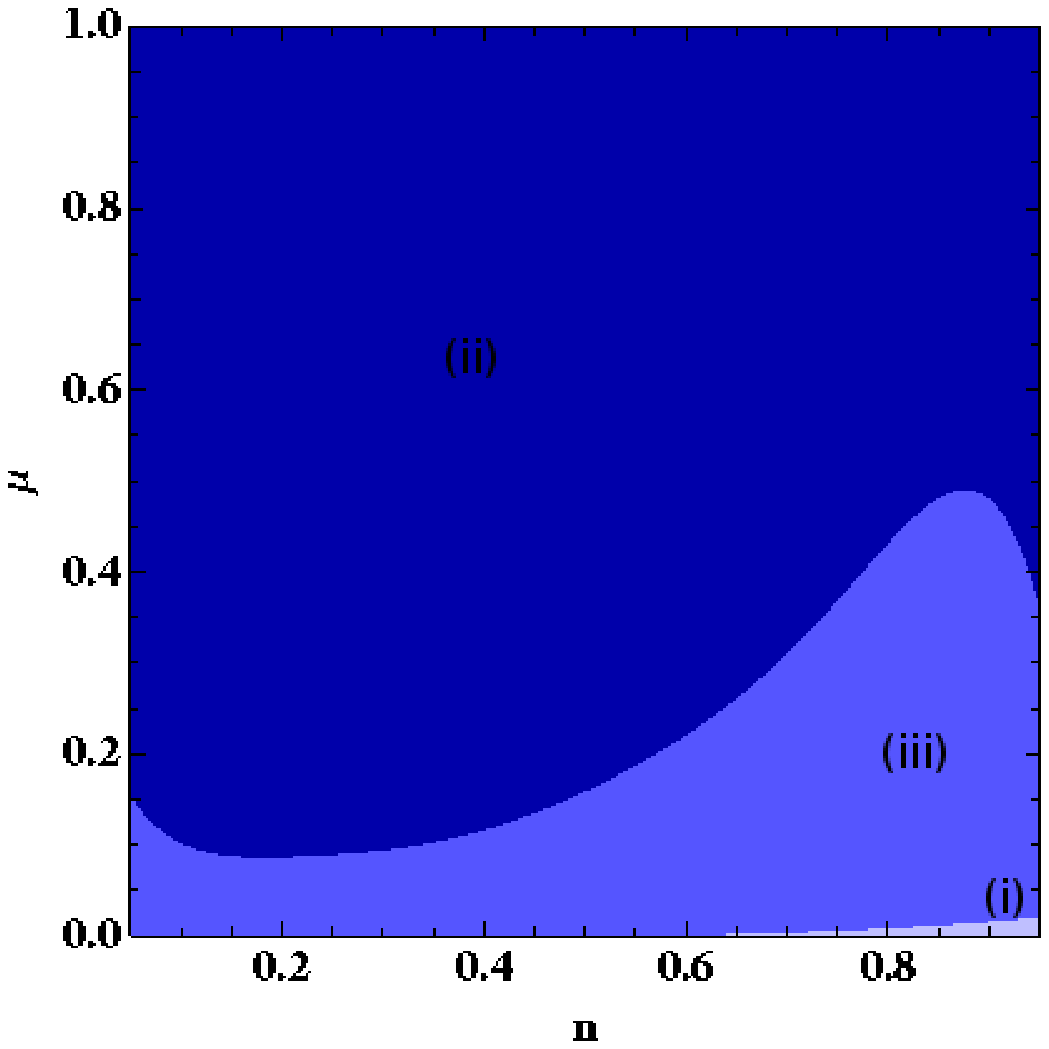}
\includegraphics[scale=.7]{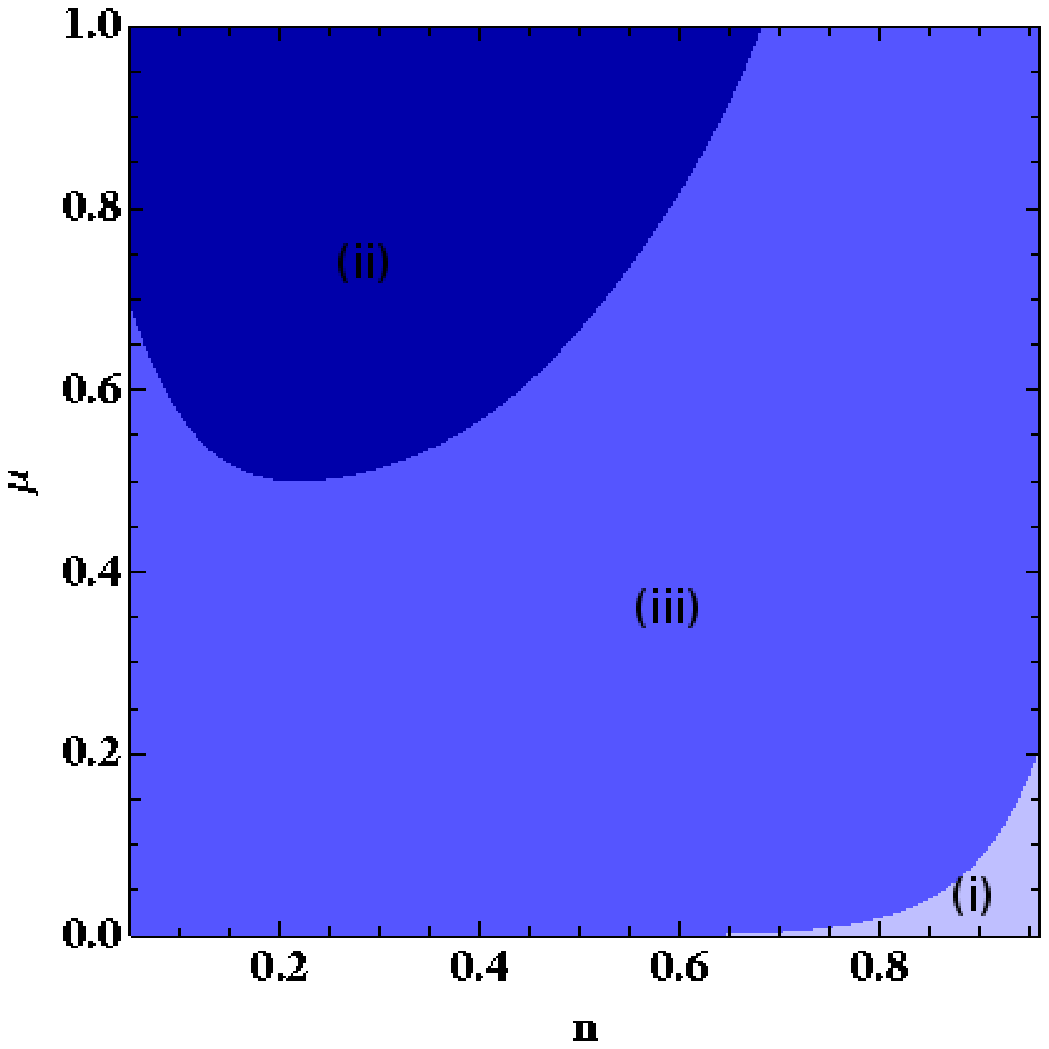}
\caption{\label{regions} 
The regions (i), (ii) and (iii) in the ($n$,$\mu$) space for $Q=1/\sqrt{6}$ (top) 
and for $Q=1$ (bottom). 
In the region (i) all the modes have $\gamma_0>0.53$.
In region (ii) $\gamma_0$ is degenerate with the value smaller than $0.5$, 
and finally (iii) shows the regime where $\gamma_0$ is dispersed.
It is clear that, for both choices of $Q$, there are viable 
choices of parameters in which the deviation from the $\Lambda$CDM 
model is present.}
\end{figure}

If we write $z_c$ explicitly in terms of the parameters and constants 
of the model, we obtain
\be
z_c=\left[ \frac{(\mu n)^{\frac{1}{1-n}}}{1-n} 
\frac{k^2M_{\rm pl}^2}{M^4}
 \left( Q \frac{\rho_m^{(0)}}{M^4}
\right)^{-\frac{2-n}{1-n}} \right]^{\frac{1-n}{4-n}}-1\,.
\label{transrshift2}
\ee
This shows that $z_c$ gets larger with increasing $\mu$ 
(as the deviation from the $\Lambda$CDM model is more significant)
and/or decreasing $Q$. 
Although the transition occurs earlier for a weaker coupling $Q$, 
we need to take into account the fact that the growth rate in the regime 
$z<z_c$ is smaller for a weaker coupling.

Let us proceed to the numerical analysis of the growth of perturbations. 
In Fig.~\ref{growth1} we plot the evolution of the growth indices 
$\gamma$ for $Q=1/\sqrt{6}$, $n=0.8$ and $\mu=0.5$ 
for a number of different wavenumbers within the range specified 
by Eq.~(\ref{krange}). 
Note that these parameters satisfy the local gravity constraints 
discussed in Sec.~\ref{lgcsec}, see Fig.~\ref{fig4}. 
At the present epoch these modes are in the regime (ii), 
with very similar growth indices today ($\gamma_0 \simeq 0.46$). 
As estimated by Eq.~(\ref{transrshift2}) the transition redshift $z_c$
is larger than the order of 1 for $k \gtrsim 0.05\,h$\,Mpc$^{-1}$, 
e.g. $z_c=2.3$ for $k=0.1\,h$\,Mpc$^{-1}$.
The degenerate behavior similar to that shown in Fig.~\ref{growth1} 
has been also found in some $f(R)$ models \cite{GMP08,Tsujikawa:2009ku}. 
Numerically we have verified that values of $\gamma_0$ in this case 
are slightly higher than those expected in the asymptotic 
regime ($\lambda  \ll \lambda_c$). 
This discrepancy comes from the fact that the chameleon 
field is lagging behind the minimum of the potential for this choice of
parameters. As a result, we need to solve the full perturbation 
equations (\ref{delphi}) and (\ref{deldeltam}) instead of the approximated
equation (\ref{delta}). Our numerical results show that 
there can be a discrepancy of up to a few percent 
in the growth rate calculated with the two different methods.

Another choice of parameters, which are compatible with 
local gravity constraints, is made in Fig.~\ref{growth2}, 
where we see the evolution of the growth indices $\gamma$ for 
$Q=1$, $n=0.7$ and $\mu=0.05$. 
The behavior of the growth indices is very different from the 
one shown in Fig.~\ref{growth1}. 
Clearly this corresponds to the regime (iii), 
in which $\gamma_0$ is dispersed with respect to 
the wavenumbers $k$.
The reason for the dispersion is that the transition
to the regime $\lambda \ll \lambda_c$ 
occurs on lower redshifts than in the case shown 
in Fig.~\ref{growth1}.

In Fig.~\ref{regions} we illustrate three different regimes 
in the $(n,\mu)$ plane for the couplings $Q=1/\sqrt{6}$ 
and $Q=1$. For the model parameters close to $(n, \mu)=(1,0)$, 
the perturbations behave similarly to those in the $\Lambda$CDM 
model [i.e., in the region (i)]. 
Figure \ref{regions} shows that the limits imposed by the constraints 
derived in Sec.~\ref{lgcsec}, although strong, allow the large parameter 
space for the existence of an enhanced growth of matter perturbations 
and for the presence of the dispersion of $\gamma_0$.
Finally we note that the large variation of $\gamma$
in the regime $z \lesssim 1$ seen in Figs.~\ref{growth1}
and \ref{growth2} will also enable us to distinguish 
the chameleon models from the $\Lambda$CDM model.

\section{Summary and Conclusions}

In this paper we have studied observational signatures 
of a chameleon scalar field coupled to non-relativistic matter.
If the chameleon field is responsible for the late-time cosmic
acceleration, the field potentials need to be consistent with
the small energy scale of dark energy as well as 
local gravity constraints.
We showed that the inverse power potential cannot satisfy 
both cosmological and local gravity constraints. 
In general, we require that the chameleon potentials are of the form
$V(\phi)=M^4 [1+f(\phi)]$, where the function $f(\phi)$ is smaller than 1 
today and $M$ is a mass that corresponds to 
the dark energy scale ($M \sim 10^{-12}$\,GeV).

The potential $V(\phi)=M^4 \exp [\mu (M/\phi)^n]$ is one of those viable
candidates. However we showed that the allowed model parameter space 
is tightly constrained by the 2006 E\"{o}t-Wash experiment.
As we see in Fig.~\ref{fig3}, the natural parameters with $n$ and $Q$ of the
order of unity are excluded for $\mu=1$.
Unless we choose unnatural values of $\mu$ smaller than $10^{-5}$, 
this potential is incompatible with local gravity constraints for 
$\{n, Q\}={\cal O}(1)$.

On the other hand, the novel chameleon potential 
$V(\phi)=M^4 [1-\mu (1-e^{-\phi/M_{\rm pl}})^n]$, 
which has the asymptotic form 
$V(\phi) \simeq M^4 [1-\mu (\phi/M_{\rm pl})^n]$ in the regime
$\phi \ll M_{\rm pl}$, can be consistent with a number of local gravity experiments
as well as cosmological constraints.
In fact this case covers the viable potentials of $f(R)$ dark energy models
in the Einstein frame. The allowed parameter regions in the $(n,Q)$ plane are illustrated in Figs.~\ref{fig4} and \ref{fig5} for $\mu=0.5$ and $\mu=0.05$. This potential is viable for natural model parameters and for the coupling $Q$ of the order of unity.

In order to distinguish the chameleon models from the $\Lambda$CDM model
at the background level, we discussed the evolution of the statefinders $(r,s)$ defined in Eq.~(\ref{sfinders}). Unlike the $\Lambda$CDM model in which $r$ and $s$ are constant ($r=1$, $s=0$) the statefinders exhibit a peculiar evolution, 
as plotted in Fig.~\ref{statefinder}. 
For the potential $V(\phi)=M^4 [1-\mu (1-e^{-\phi/M_{\rm pl}})^n]$ we found 
that $r<1$ and $s>0$ around the present epoch, but the solutions approach 
the de Sitter point $(r,s)=(1,0)$ in future. The upcoming observations of SN Ia may discriminate such an evolution from other dark energy models. 

We have also studied the growth of matter perturbations for the chameleon 
potential $V(\phi)=M^4 [1-\mu (1-e^{-\phi/M_{\rm pl}})^n]$. 
The presence of a fifth force between the field 
and non-relativistic matter (dark matter/baryons) modifies the equation
of matter perturbations, provided that the field mass $m_{\phi}$ is smaller 
than the physical wavenumber $k/a$, or $\lambda < \lambda_c$. 
Cosmologically the field is heavy in the past (i.e. for large density), 
but the mass $m_{\phi}$ decreases by today (typically of the order of $H_0$)
in order to realize the late-time cosmic acceleration.
Then the transition from the regime $k/a<m_{\phi}$ ($\lambda > \lambda_c$) 
to the regime $k/a>m_{\phi}$ ($\lambda < \lambda_c$) can occur at the redshift 
$z_c$ given in Eq.~(\ref{transrshift2}). For the perturbations on smaller scales 
(i.e. larger $k$) the critical redshift $z_c$ tends to be larger. 

For the model parameters and the coupling $Q$ bounded by 
a number of experimental and cosmological constraints, we have 
studied the evolution of the growth index $\gamma$ of matter 
perturbations. Apart from the ``General relativistic regime'' in 
which two parameters $n$ and $\mu$ of the potential (\ref{vipo2}) 
are close to $(n,\mu)=(1,0)$, we found that the values 
of $\gamma$ today exhibit either dispersion with respect 
to the wavenumbers $k$ (region (iii) in Fig.~\ref{regions}) or no 
dispersion, however with $\gamma_0$ smaller than 0.5 (region (ii) 
in Fig.~\ref{regions}). Both cases can be distinguished from the $\Lambda$CDM model 
(where $\gamma \simeq 0.55$). Moreover, as seen in Figs.~\ref{growth1} 
and \ref{growth2}, the variation of $\gamma$ on low redshifts is significant. 

From observations of galaxy clustering we have not yet obtained 
the accurate evolution of $\gamma$. This is linked to the fact that all probes of clustering are plagued by a bias problem. However upcoming galaxy surveys may pin down the matter power spectrum to exquisite accuracy,
together with a better understanding of bias. In order to confirm or 
rule out models like ours, one must also address the observability of $\gamma$ 
both as a function of $z$ and $k$. We hope that future observations will 
provide an exciting possibility to detect the fifth force induced by the 
chameleon scalar field.

\section*{Acknowledgments}
DP thanks JSPS for financial support during his stay at Tokyo University 
of Science. 
DFM and HAW thanks the Research Council of Norway FRINAT grant 197251/V30. DFM is
also partially supported by project CERN/FP/109381/2009 and
PTDC/FIS/102742/2008. BM thanks Research Council of Norway (Yggdrasil Program grant No. 202629V11) 
for financial support during his stay at the University of Oslo where part of this 
work was carried out and thanks DFM and HAW for the hospitality. 
RG thanks CTP, Jamia Millia Islamia for hospitality where
a part of this work was carried out.
The work of ST was supported by the Grant-in-Aid for Scientific Research 
Fund of the JSPS No.~30318802 and by the Grant-in-Aid for
Scientific Research on Innovative Areas (No.~21111006). 
ST thanks Savvas Nesseris, Kazuya Koyama, 
Burin Gumjudpai, and Jungjai Lee for warm hospitalities 
during his stays in the Niels Bohr Institute, the University of Portsmouth, 
Naresuan University, and Daejeon.


\end{document}